 \documentclass[twocolumn,amsmath,amssymb]{article}
    
\usepackage{dcolumn}
\usepackage{bm}
\usepackage{hyperref}
\usepackage{amsmath} 
\usepackage{amsfonts}  
\usepackage{graphics}   
\usepackage{graphicx} 
\usepackage{epsfig}    
\usepackage{rotating}  
 \usepackage[latin1]{inputenc}
 \usepackage{color}
 \usepackage{rotating}
 \usepackage{setspace}
\usepackage{natbib}

\begin{document}

\twocolumn[    

\begin{center}
{\Large \bf
 Self-Organization of Mobile Populations in Cyclic Competition}\\

\vspace*{0.5cm}

{\large Tobias Reichenbach $^{1}$,  Mauro Mobilia $^{1,2}$, and Erwin Frey $^{1}$}\\

\vspace*{0.5cm}

{\small $^{1}$Arnold Sommerfeld Center for Theoretical Physics and  Center for NanoScience, Department of Physics, Ludwig-Maximilians-Universit\"at M\"unchen, Theresienstrasse 37, D-80333 M\"unchen, Germany.\\
$^{2}$Mathematics Institute and Warwick Centre
for Complexity Science, The University of Warwick, Gibbet Hill Road, Coventry CV4 7AL, United Kingdom
}\\

\end{center}

\vspace*{0.5cm}

{\bf Abstract}\\

 The formation of out-of-equilibrium
patterns is a characteristic feature of spatially-extended, biodiverse, ecological systems. Intriguing examples are provided by cyclic competition of species, as metaphorically described by the `rock-paper-scissors' game. Both experimentally and theoretically, such non-transitive interactions  have been found to induce self-organization of static individuals into noisy, irregular clusters. However, a profound understanding and characterization of such patterns is still lacking. 
Here,  we theoretically investigate the influence of individuals' mobility on the spatial structures emerging in rock-paper-scissors games. We
 devise a quantitative approach to analyze the spatial patterns self-forming in the course of the
stochastic time evolution. For a paradigmatic model originally introduced by May and Leonard, 
within an interacting particle approach, we demonstrate that the system's behavior - in the proper continuum limit -
is aptly captured by a set of stochastic partial differential equations. 
The system's stochastic dynamics is shown to lead to the emergence of  entangled rotating spiral waves. 
While the spirals' wavelength and spreading velocity is demonstrated to be accurately predicted by a (deterministic) complex Ginzburg-Landau equation, their entanglement results from the inherent stochastic nature of the system. 
These findings and our methods have important applications for understanding the formation of noisy patterns, e.g., in  ecological and evolutionary contexts, and are also of relevance for the kinetics of (bio)-chemical reactions.\\
\vspace*{0.3cm}

]

\section{Introduction}

Spatial distribution of individuals, as well as their mobility, are  common features of real ecosystems that often  come paired~\citep{May}. On all scales of living organisms, from bacteria residing in soil or on Petri dishes, to the largest animals living in savannas - like elephants - or in forests, populations' habitats are spatially extended and individuals interact locally within their neighborhood. Field studies as well as experimental and theoretical investigations have shown that the locality of the interactions
leads to the self-formation of complex spatial patterns~\citep{May,Murray,turing-1952-237,nowak-1992-359,hassell-1991-353,hassell-1994-370,blasius-1999-399,kerr-2002-418,king-2003-64, hauert-2004-428, scanlon-2007-449, kefi-2007-449,szabo-2007-446,perc-2007-75, Nowak}.  Another important property of most individuals is mobility. For example,  bacteria swim and tumble, and animals migrate. As motile individuals are capable of enlarging their district of residence, mobility may  be viewed as a mixing, or stirring mechanism which ``counteracts'' the locality of spatial interactions.

\begin{figure*}  
\begin{center}  
\includegraphics[width=12cm]{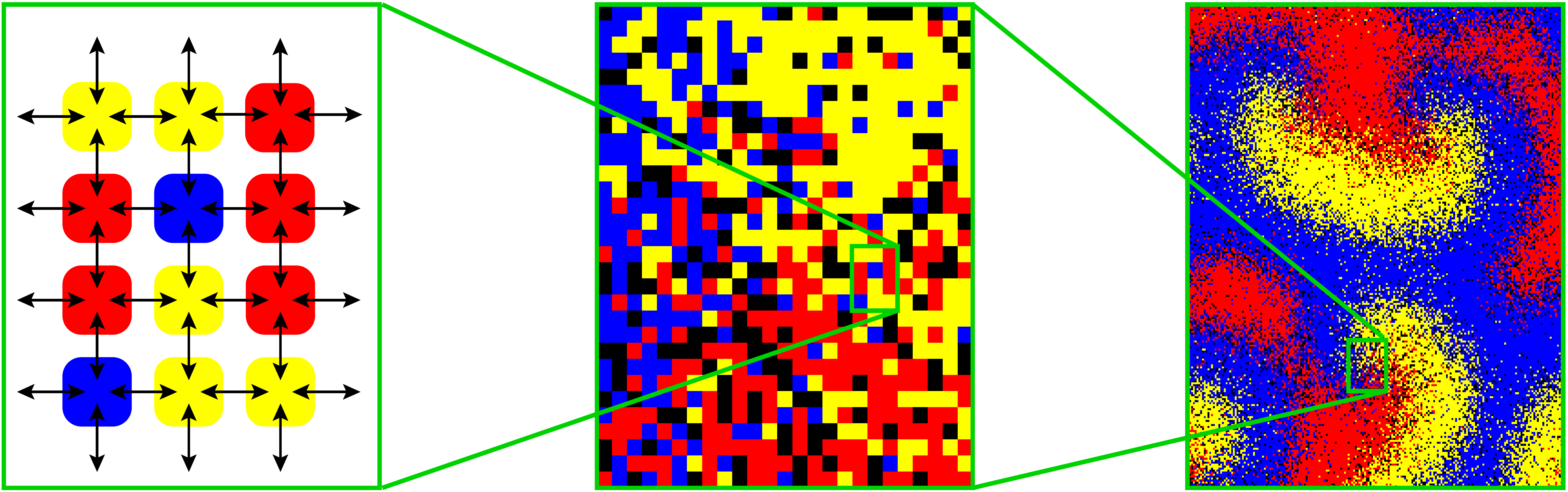}
\caption{\small The stochastic spatial system at different scales.
Here, each of the colors yellow, red, and blue (level of gray) represents one species, and black dots identify empty spots.
 Left: Individuals are arranged on a spatial lattice and randomly interact with their nearest neighbors. Middle: At the scale of about 1,000 individuals, stochastic effects dominate the system's appearance, although domains dominated by different subpopulations can already be detected. Right: About 50,000 mobile interacting individuals self-organize into surprisingly regular spiral waves.  
\label{scales}}  
\end{center}                
\end{figure*}

The combined influence of these effects, i.e. the competition between mobility and spatial separation,  
on the spatio-temporal development of populations is one of the most interesting and complex problems in theoretical ecology~\citep{May,Murray,turing-1952-237,hassell-1994-370,king-2003-64,janssen-2001-103,reichenbach-2007-448}. If mobility is low, locally interacting populations can exhibit involved spatio-temporal patterns, like traveling waves~\citep{igoshin-2004-101}, and for example lead to the self-organization of individuals into spirals in myxobacteria aggregation \citep{igoshin-2004-101} and insect host-parasitoid populations~\citep{hassell-1991-353}, or more fractal-like structures in competing strains of \emph{E.coli} \citep{kerr-2002-418}. On the other hand, high  mobility results in  well-mixed systems where the spatial distribution of the populations is irrelevant \citep{Smith,Hofbauer}. In this situation, spatial patterns do no longer form: The system adopts a spatially uniform state, which therefore drastically differs from the low-mobility scenario.

An intriguing motif of the complex competitions in a population, promoting species diversity, is constituted by  three subpopulations exhibiting cyclic dominance. This basic motif is metaphorically described by the rock-paper-scissors game, where rock crushes scissors, scissors cut paper, and paper wraps rock. Such non-hierarchical, cyclic competitions, where each species outperforms another, but is also itself outperformed by a remaining one, have been identified in different ecosystems like coral reef invertebrates \citep{jackson-1975-72}, rodents in the high-Arctic tundra in Greenland \citep{gilg-2003-302}, lizards in the inner Coast Range of California \citep{sinervo-1996-340}  and  microbial populations of colicinogenic \emph{E. coli} \citep{kerr-2002-418,kirkup-2004-428}. In the latter situation, it has been shown that spatial arrangement of quasi-immobile bacteria (because of `hard'  nutrient or substrate)   on a Petri-dish leads to the stable coexistence of all three competing bacterial strains, with the formation of irregular patterns. In stark contrast, when the system is well-mixed, there is spatial homogeneity resulting in the take over of one subpopulation and the extinction of the others after a short transient.

It is worth noting that the emergence of noisy patterns, as those studied here, is a feature shared across disciplines by many complex systems characterized by their out-of-equilibrium nature and nonlinear interactions. Examples range from the celebrated  Belousov-Zhabotinsky reaction \citep{zaikin-1970-225} (spiralling patterns) and many other chemical reactions \citep{Kapral}, to epidemic outbreaks (traveling waves) \citep{grenfell-2001-414,cummings-2004-427}, excitable media \citep{muratov-2007-104,Kapral}, and calcium signalling within single cells \citep{lechleiter-1991-252,falcke-2004-53,bootmann-2006-119}. Moreover, cyclic dynamics as described by the rock-paper-scissors game occur not only in population dynamics but have, e.g., been observed in social dilemmas relevant in behavioral sciences~\citep{sigmund-2001-98,hauert-2002-296}.
Therefore, we would like to emphasize that the methods presented in this work 
are not limited to theoretical ecology and biology, but have a broad range  of
multidisciplinary applications and notably include the above fields.

Pioneering work on the role of mobility in ecosystems was performed by~\citet{levin-1974-108}, where the dynamics of a population residing in two coupled patches was investigated: Within a deterministic description, Levin identified a critical  value for the individuals' mobility
between the patches. Below the critical threshold, all subpopulations coexisted, while only one remained above that value. Later, more realistic models of many patches, partly spatially arranged, were also studied, see~\citet{hassell-1991-353, hassell-1994-370,blasius-1999-399,alonso-2002-64} as well as references therein. These works shed light on the formation of patterns, in particular traveling waves and spirals. However, patch models have been criticized 
for treating the space in an ``implicit'' manner (i.e. in the form of coupled habitats without internal structure)~\citep{durrett-1998-53}. In addition, the above investigations were often restricted to deterministic 
dynamics and thus did not address the spatio-temporal influence of noise. To overcome these limitations, \citet{durrett-1997-185} 
 proposed to consider interacting particle systems, i.e.
stochastic spatial models with populations of discrete individuals distributed on lattices. In this realm, studies   have mainly focused on numerical simulations and on  (often heuristic) deterministic reaction-diffusion equations, or coupled maps \citep{durrett-1994-46,durrett-1997-185,durrett-1998-53,king-2003-64,czaran-2002-99,nowak-2005-433,mobilia-2006-73,mobilia-2007-128, szabo-2007-446}. 

Here, we demonstrate how a - spatially explicit -  stochastic model of cyclically interacting subpopulations exhibits self-formation of spatial structures
which, in the presence of individuals' mobility,  turn into surprisingly regular, geometric spiral waves. The latter become visible on the scale of a large number of interacting individuals, see  Fig.~\ref{scales} (right).  In contrast, stochastic effects solely dominate on the scale of a few individuals, see Fig.~\ref{scales} (left), which interact locally with their nearest neighbors. Spatial separation of subpopulations starts to form on an intermediate scale, Fig.~\ref{scales} (middle), where mobility leads to fuzzy domain boundaries, with major contributions of noise. On a larger scale,  Fig.~\ref{scales} (right), these fuzzy patterns adopt regular geometric shapes. As shown below, the latter are jointly determined by the deterministic dynamics and intrinsic stochastic effects. 
In the following, we elucidate this subtle interplay by mapping - in the continuum limit - the stochastic spatial dynamics onto a set of stochastic partial differential equations (SPDE) and, using tools of dynamical
systems (such as normal forms and invariant manifolds), by recasting the 
underlying deterministic kinetics in the form of a complex Ginzburg-Landau equation (CGLE). The CGLE allows 
us to make analytical predictions for the spreading velocity and wavelength
of the emerging spirals waves. 
Below, we provide a detailed description of 
these methods and convey a thorough discussion of the spatio-temporal properties of the system with an emphasis on the role of spatial degrees of freedom, mobility and internal noise.

In our first article on this subject~\citep{reichenbach-2007-448} we have described how a mobility threshold separates a biodiverse regime (arising for low mobilities) from a high-mobility regime where diversity is rapidly lost. In~\citep{reichenbach-2007-99} we have further analyzed the travelling spiral waves that arise for low mobilities  and computed correlation functions as well as the spirals' wavelength and spreading velocity. In this article, we provide
a comprehensive discussion of the quantitative analysis of the system's properties. This includes the detailed derivation  of all mathematical equations,  an accurate description of the numerical simulations (via the implementation of an efficient algorithm for the lattice simulations taking exchange processes into account) as well as the analytical treatment of the out-of-equilibrium patterns  emerging in the course of the time evolution.

\section{Simulations and Results}

We study a stochastic spatially-extended version of a three species model originally investigated (on rate equations level) by~\citet{may-1975-29}.
In~\citep{reichenbach-2007-448} we have already considered some properties of such a model and demonstrated the existence of  a critical value of the populations' mobility separating a biodiverse regime, where all subpopulations coexist, from a uniform regime, where only one subpopulation survives. A brief account of the analysis of the spatio-temporal properties of the coexistence phase has recently been given in~\citep{reichenbach-2007-99}. Here, we complete and considerably extend those previous works by giving a comprehensive picture of the system's properties and details of various mathematical methods (interacting particle systems, stochastic processes, dynamical systems, partial differential equations) allowing to deal with these kinds of multidisciplinary problems.

\begin{figure}    
\begin{center} 
\includegraphics[scale=1]{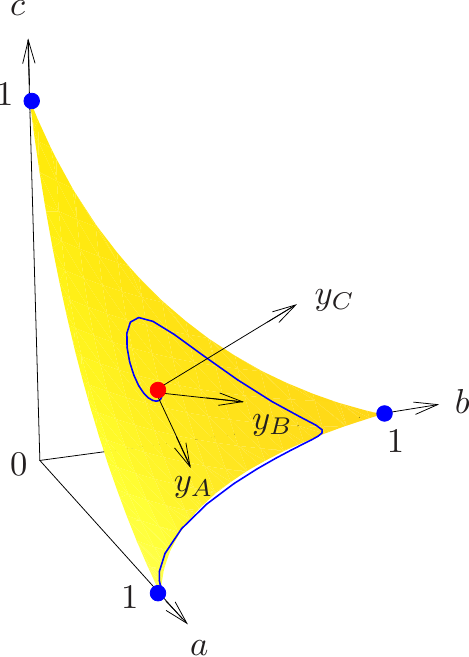}
\caption{\small The phase space of the non-spatial system. It is spanned by the densities $a,b$, and $c$ of species $A$, $B$, and $C$. 
On an invariant manifold (in yellow/light gray, see text), the flows obtained as solutions of the rate equations~(\ref{ml_rate_eqs}) (an example is shown in blue/dark gray) initially in the vicinity of the reactive fixed point (red/gray, see text) spiral outwards, approaching the heteroclinic cycle which connects three trivial fixed points (blue/dark gray). In Subsection~\ref{subsec:IM}, we introduce the appropriate coordinates $(y_A,y_B,y_C)$  which reveal the mathematical structure of the manifold and reflect the cyclic symmetry of the system.
\label{pict_coord}}  
\end{center}
\end{figure}
Consider three subpopulations $A,~B$ and $C$ which cyclically dominate each other. An individual of subpopulation $A$ outperforms a $B$ individual through ``killing'' (or ``consuming''), symbolized by the (``chemical'') reaction  $AB\rightarrow A\oslash$, where $\oslash$ denotes an available empty space. In the same way, $B$ outperforms $C$, and $C$ beats $A$ in turn, closing the cycle. We refer to these processes as selection and denote the corresponding rate by $\sigma$. To mimic a finite carrying capacity, we allow each subpopulation to reproduce only if an empty space is available, as described by the reaction $A\oslash\rightarrow AA$ and analogously for $B$ and $C$. For all subpopulations, these reproduction events occur with rate $\mu$, such that the three subpopulations equally compete for empty space. To summarize, the reactions that define the model (selection and reproduction) read 
\begin{eqnarray}
AB&\stackrel{\sigma}{\longrightarrow} A\oslash\,, &\qquad A\oslash \stackrel{\mu}{\longrightarrow}AA\,, \cr
BC&\stackrel{\sigma}{\longrightarrow} B\oslash\,, &\qquad B\oslash \stackrel{\mu}{\longrightarrow}BB\,,\cr
CA&\stackrel{\sigma}{\longrightarrow} C\oslash\,, &\qquad C\oslash \stackrel{\mu}{\longrightarrow}CC\,.
\label{ml_react}
\end{eqnarray}  
In the absence of spatial degrees of freedom, the kinetics of these 
reactions are embodied by rate equations for the temporal evolution of the mean densities $a(t), b(t), c(t)$ of the subpopulations $A, B$ and $C$, respectively, given by Eqs.~(\ref{ml_rate_eqs}) in Subsection~\ref{subsec:RE}. These equations provide a deterministic description which is well suited to describe  well-mixed systems with a large number of individuals, such as Moran processes \citep{moran-1958-54,traulsen-2005-95,traulsen-2006-74} or urn models \citep{Feller,reichenbach-2006-74}. For the system under consideration, the rate equations are given and carefully investigated in Subsection~\ref{subsec:RE}. As main features, they possess three absorbing fixed points corresponding to survival of only one subpopulation (the solution corresponding to an empty system is also an absorbing state, but is irrelevant for our purpose and will be ignored thereafter), as well as a reactive fixed point where all three subpopulations coexist, see Fig.~\ref{pict_coord}. The coexistence fixed point is unstable, and trajectories starting in its vicinity spiral outwards. The spiralling flows lie on an invariant manifold, and approach a heteroclinic cycle which connects the three absorbing fixed points. The approaching trajectories spend longer and longer periods of time in a neighborhood of each (absorbing) fixed points, before departing to the next one. This means that the system alternates between states where nearly only one of the three species is present, with rapidly increasing time period. However, this picture is idealized as it crucially (and tacitly) assumes the presence of an infinite population. In fact, fluctuations,  e.g. stemming from a finite number of individuals, lead to reach one of the absorbing states where only one subpopulation takes over the whole system. Which of the absorbing states is reached depends on the initial conditions as well as on random fluctuations. Due to the symmetry of the reactions~(\ref{ml_react}), when averaging over all possible initial conditions as well as fluctuations, all species have an equal chance to survive.

In the spatially-extended stochastic version of the model, we adopt an interacting particle description where individuals of all subpopulations are arranged on a lattice. In this approach, each site of the grid is either occupied by one individual or empty, meaning that the system has a finite carrying capacity, and the reactions~(\ref{ml_react}) are then only allowed between \emph{nearest neighbors}. In addition, we endow the individuals with a certain form of mobility. Namely, at rate $\epsilon$ all individuals can exchange their position with a nearest neighbor. With that same rate $\epsilon$, any individual can also hop on a neighboring empty site. These exchange processes lead to an effective diffusion of the individuals.
Denote $L$ the linear size of the $d$-dimensional hypercubic lattice (i.e. the number of sites along one edge), such that the total number of sites reads $N=L^d$. Choosing the linear dimension of the lattice as the basic length unit, the macroscopic diffusion constant $D$ of individuals stemming from exchange processes reads 
\begin{equation}
D=\epsilon d^{-1} N^{-2/d}\,;
\label{eps_scaling}
\end{equation}
the derivation of this relation is detailed in Subsection~\ref{subsec:cont_limit}.

How do individual's mobility  and internal noise, in addition to nonlinearity, affect the system's behavior ?
\\
Insight into this ecologically important issue can be gained from a continuum limit where the diffusion constant
$D$ is \emph{finite}. Namely, we investigate the limit of infinite system size, $N\to\infty$, where the diffusion $D$ is kept constant (implying that the local exchange rate tends to infinity, $\epsilon\to \infty$). In Subsections~\ref{subsec:cont_limit} and~\ref{subsec:noise}, we show how, in this limit, a description of the stochastic lattice system through stochastic partial differential equations (SPDE) becomes feasible. These SPDE describe the time evolution of the (spatially dependent) densities 
$a(\bm r,t), b(\bm r,t), c(\bm r,t)$ of the subpopulations $A, B$, and $C$, respectively. The expression of the SPDE, Eqs.~(\ref{stoch_part_eq}), is given in Subsection~\ref{subsec:noise} along with their detailed derivation. The latter relies on a system-size expansion in the continuum limit which allows to obtain the noise terms of the SPDE. Noise stems from the stochasticity of the reactions~(\ref{ml_react}) as well as from the discreteness and the finite number of individuals.

To investigate the behavior of the interacting particle system and to compare it with the predictions of the  SPDE~(\ref{stoch_part_eq}), we have carried out stochastic simulations of the model on a square lattice with periodic boundary conditions and of size ranging from $L=50$ up to $L=1000$ sites. In the following, we always consider the system in two spatial dimensions, $d=2$. At each simulation step, a randomly chosen individual interacts with one of its four nearest neighbors, being also randomly determined. In a straightforward  algorithm,  at each Monte Carlo (MC) step, one determines (via a random number) which reaction (exchange, selection, or reproduction) occurs next. Reproduction happens with probability $\mu/(\mu+\sigma+\epsilon)$, selection with $\sigma/(\mu+\sigma+\epsilon)$, and exchange events occur with probability $\epsilon/(\mu+\sigma+\epsilon)$. Then, a random pair of nearest neighbors is selected and the chosen type of interaction (reproduction, selection or exchange) is performed, if the move is allowed. In our situation, a more efficient algorithm inspired by~\citet{gillespie-1976-22,gillespie-1977-81} can be implemented.  Namely, in the continuum limit $N\to \infty$, the exchange rate $\epsilon$ becomes large compared to the selection and reproduction rates, $\mu$ and $\sigma$.  
Thus, a  large number of exchange events  occurs between two reactions~(\ref{ml_react}). Indeed, the probability $P$ of having $E$ exchanges between two subsequent May-Leonard reactions reads
\begin{equation}
\label{P_E}
P(E)=\left(\frac{\epsilon}{\mu+\sigma+\epsilon}\right)^E\left(\frac{\mu+\sigma}{\mu+\sigma+\epsilon}\right).
\end{equation}  
Hereby, the first factor on the right hand side denotes the probability of subsequently drawing $E$ exchange events, and the second factor is the probability that the next event is either a selection or reproduction process. 
To efficiently take this high number of exchanges occurring between selection/reproduction processes into account,
at each MC step, we draw the number of such exchange events from the probability distribution (\ref{P_E}). 
This number of random exchanges is performed, and then followed by one of the May-Leonard reactions~(\ref{ml_react}).

All results we present from lattice simulations were obtained starting from 
a random initial distribution of individuals and vacancies: the probability for each site to be in one of the four possible states (i.e. $A,B, C$ or $\oslash$) has been chosen to coincide with the  value
of the (unstable) internal fixed point of the rate equations~(\ref{ml_rate_eqs}). Thereafter, without loss of generality (see Section 3.6), we often consider equal selection and reproduction rates, which we set to one  (thereby defining the time scale), i.e. $\mu=\sigma=1$. In this case, all four states initially occur with equal probability $1/4$.  Generally,  after a short transient, a reactive steady state emerges.
Hereby, the discussion of the stability of the reactive steady states should be dealt with some care because the only 
absorbing states are those where the  system is uniformly covered by only one subpopulation. Reactive states 
are not stable in a strict sense because they can be spoilt by chance fluctuations (the system is large but {\it finite})
which drive the system into one of the absorbing uniform states. 
However, the typical (average) waiting time $T$ to reach the extinction of two subpopulations  is extremely long for large systems and it diverges with increasing system size~\citep{reichenbach-2007-448}, implying the existence of  super-persistent transients~\citep{hastings-2004-19}. To rationalize this point, here we follow~\citep{reichenbach-2007-448} where we have proposed to characterize the stability of the reactive steady states by comparison with the average extinction time obtained for the marginally stable version of the system, where $T\propto N$~\citep{reichenbach-2006-74}. Therefore, when the extinction time grows faster than linearly with the system size, the reactive steady state is deemed to be stable on ecologically reasonable time scales.

In the continuum limit, the system is described in terms of the SPDE~(\ref{stoch_part_eq}),
characterized by white noise of strength $\propto N^{-1/2}$ (see Section 3.5), 
 which are derived from the master equation via a system-size expansion  as explained in Appendix D. 
To compare this approach with the results of the lattice system, we have numerically solved Eqs.~(\ref{stoch_part_eq}) using open software from the XMDS project \citep{collecutt-2001-142,xmds}. For specificity, we have used spatial meshes of $200\times 200$ to $500\times 500$ points, and 10,000 steps in the time-direction. Initially, we started with homogeneous densities, corresponding to the values of the unstable internal fixed point of the RE~(\ref{ml_rate_eqs}), as in the lattice simulations. It is worthwhile noticing that noise terms of the SPDE have to be treated with special care as they may occasionally lead to the nonphysical situation of negative densities (or densities exceeding the maximal value of $1$) if the system is close to the phase space boundaries. This problem is most pronounced when  additive noise is used, while in our case, noise is multiplicative and vanishes near the boundaries.  Still, due to discretization effects, nonphysical situations of negative densities may arise; in our simulations, we have discarded such rare events.

\subsection{Spiral structures in the continuum limit}

\begin{figure}   
\begin{center}
\includegraphics[scale=1]{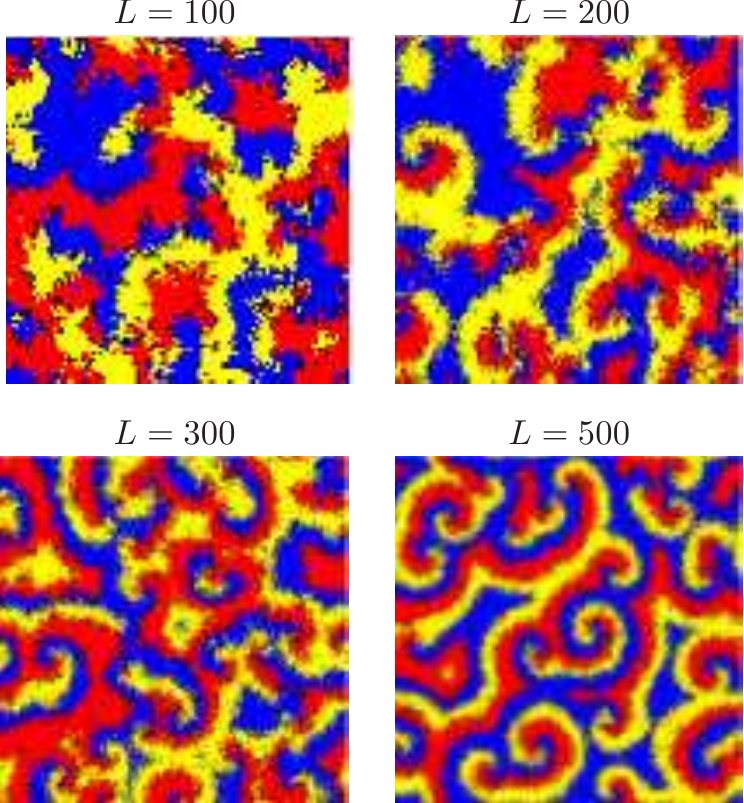}
\caption{\small Approach of the continuum limit. We show snapshots of the reactive steady state of the stochastic system, for $D=1\times 10^{-5},~\mu=\sigma=1$, and different system sizes.  Each color (level of gray) represents a different species (black dots denote empty spots).
The lattice sizes are $L=100$ ($\epsilon=0.2$) in the top left panel, $L=200$ ($\epsilon=0.8$, top right), $L=300$ ($\epsilon=1.8$, bottom left), and $L=500$ ($\epsilon=5$, bottom right). 
Increasing the system size ($D$ is kept fixed), the
continuum limit is approached.
Random patterns appear for small systems ($L=100$), while entangled spiral waves emerge when $L$ is raised 
and clearly emerge
in large systems ($L=500$).
\label{snapshots_cont}}
\end{center}                
\end{figure}
To study the system's behavior in the approach of the continuum limit, we have kept the diffusion fixed at a value  $D=1\times 10^{-5}$, and systematically varied the system size $N$ (and henceforth the exchange rate $\epsilon$, which reads $\epsilon=2DN$ in two spatial dimensions). 
In Fig.~\ref{snapshots_cont}, we report typical long-time snapshots of the reactive steady states for various values of the exchange rate and different system sizes. In small lattices, e.g. $L=100$, we observe that all  subpopulations coexist and form clustering patterns 
 of characteristic size. The spatio-temporal properties of the latter do  not  admit simple  description and appear to be essentially dominated by stochastic fluctuations. 
When the size of the lattice is large, e.g. $L=300-500$, the  three populations also coexist in a reactive steady state. However, in stark contrast from the above scenario, we now find that individuals  self-organize in an entanglement of rotating spiral waves. The properties of these spirals, such as  their wavelength, their frequency and the spreading velocity,  are  remarkably  regular
and will be studied below.

\begin{figure}
\begin{center}   
\includegraphics[scale=1]{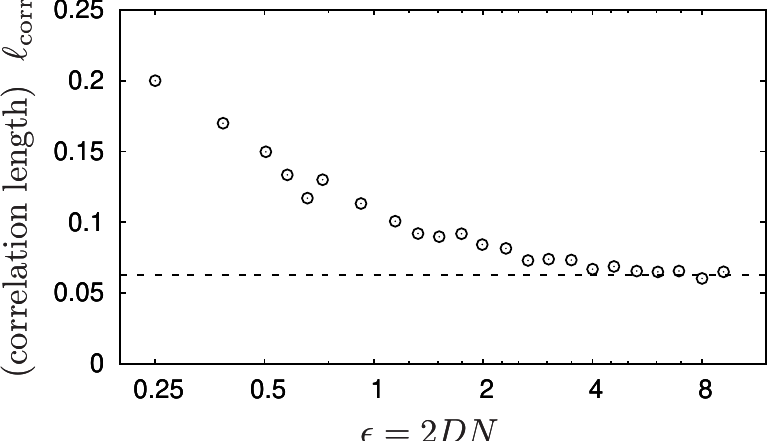}
\caption{\small The typical size of the patterns, as  described by the correlation length $\ell_\text{corr}$. 
We show its dependence on the exchange rate $\epsilon$, for fixed $D=5\times 10^{-5},~\mu=\sigma=1$, and different system sizes $N$. For large $N$, i.e. large $\epsilon$, we expect the system to be well described by SPDE, the correlation length of the latter is depicted as a dashed line. Surprisingly, the correlation length already agrees excellently with this continuum model for $\epsilon \geq 5$.
\label{eps_corr}} 
\end{center}                
\end{figure}
\begin{figure*}    
\begin{center}
\includegraphics[scale=1]{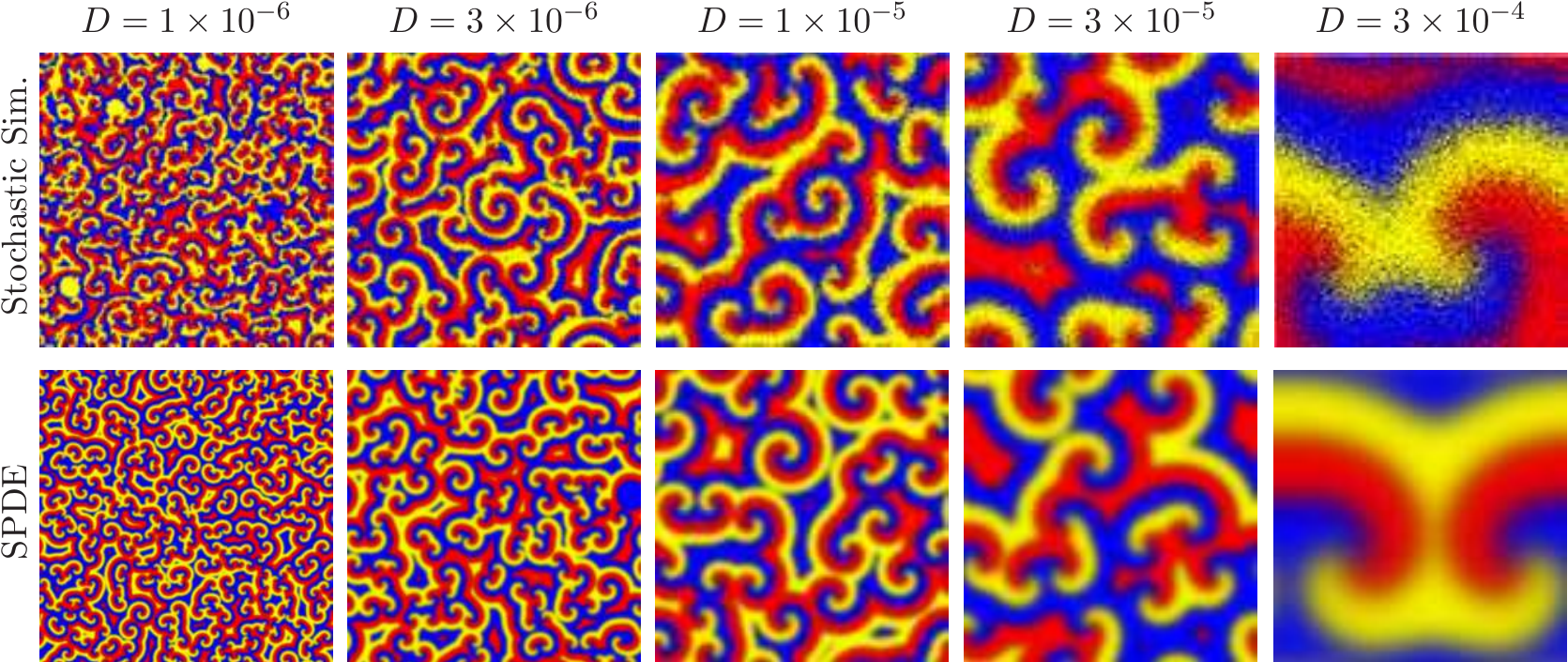}
\caption{\small
The reactive steady states. We show snapshots emerging in simulations of the interacting particle system~(\ref{ml_react}) (top row) and obtained by solving the SPDE~(\ref{stoch_part_eq}) (bottom row). 
 Each color (level of gray) represents a different species (black dots denote empty spots).
From left to right, the diffusion constant is increased from $D=1\times 10^{-6}$ to $D=3\times 10^{-4}$. The latter value is slightly below the critical threshold  above which the spiral structures can no longer fit within the system \citep{reichenbach-2007-448}; see text. The system sizes used in the stochastic simulations are $L=1000$ in the left two panels,  $L=300$ for that at right, and $L=500$ for the other two (middle). The selection and reproduction rates are chosen as $\sigma=\mu=1$. 
\label{snapshots}} 
\end{center}                
\end{figure*}
The asymptotic approach towards the continuum limit, as illustrated in the  snapshots of Fig.~\ref{snapshots_cont}, can be rationalized by considering the typical size of the patterns.
 The latter are obtained from the computation of spatial correlation functions (see next Subsection for the definition of correlation functions) and the ensuing 
correlation length $\ell_\text{corr}$, which is the length at which the correlations decay by a factor $1/e$ from their maximal value). This quantity gives the average spatial extension of the patterns. In Fig.~\ref{eps_corr}, we report $\ell_\text{corr}$ as obtained for systems of different sizes, i.e. for various values of $\epsilon$. For small systems (low exchange rate $\epsilon$), the pattern size is considerably larger than in the continuum limit (dashed line obtained from SPDE, see below). Increasing the system size (larger values of $\epsilon$), the continuum limit is approached. Remarkably, there is a striking agreement between the results from the lattice simulations and the SPDE  
already  for $\epsilon\geq 5$. Hereby, $\mu=\sigma=1$, such that the continuum limit is already closely approached when $\epsilon$ is of the same order as the rates for the selection and reproduction events, yet larger. This result is also apparent in Fig.~\ref{snapshots_cont}, where for $L=500$ and $\epsilon=5$, the system already exhibits regular spiral waves. It follows from this discussion that the
results obtained in the
continuum limit (derived assuming $\epsilon \to \infty$), actually have a broader range of validity and allow to aptly describe the interacting particle system already for  $\epsilon$ finite and of the same order of (yet larger than) $\mu$ and $\sigma$. A comparable influence of short-range mixing  has
also been reported recently for a predator-prey - or host-pathogen - model, where a short-range exchange process with finite rate has been shown to crucially affect the fate of the system (absorbing or coexistence state) through a (first-order) phase transition~\citep{mobilia-2006-73}. Furthermore, the smooth domain boundaries caused by mixing and the emerging spiral waves are similar to the spatial patterns investigated in~\citep{szabo-2002-65} that arise from  slow cyclic dynamics combined with Potts energy. The authors of this study have analyzed the resulting spirals by considering the vortex density and average length of vortex edges.

In Fig.~\ref{snapshots}, we report snapshots of the reactive steady state obtained from the stochastic simulations (left column) and as predicted by the SPDE (right column)  for different values of the diffusion constant $D$. In Fig.~\ref{snapshots}, panels in the same row have been  obtained for the same parameters (for lattices of size $L=300,500$ and $1000$). We observe an excellent qualitative and quantitative agreement between both descriptions, which yield the formation of rotating spiral waves whose typical sizes and wavelengths manifestly coincide (see  below). \underline{}
When the diffusion constant $D$ is increased,  the size of the spirals increases, too. With help of the underlying SPDE~(\ref{stoch_part_eq}), this observation can be rationalized by noting that the size of the spatial structures is proportional to $\sqrt{D}$. 
This scaling relation stems from the fact that spatial degrees of freedom only enter the Eqs.~(\ref{stoch_part_eq}) through the diffusion term $D\Delta$, where the Laplacian operator $\Delta$ involves second-order spatial derivatives. 
Therefore, rescaling the spatial coordinates by a factor $1/\sqrt{D}$ makes  the diffusive contribution independent of $D$, with the ensuing scaling relation. As we have shown in~\citep{reichenbach-2007-448}, the three subpopulations can coexist only up to a critical value of the diffusion rate. Above that threshold, the spirals outgrow the system and there is extinction of two populations: only one subpopulation (at random) survives and covers the entire lattice.

\subsection{Correlations}

The comparison of snapshots obtained from lattice simulations with the numerical solutions of the SPDE  reveals a remarkable coincidence of both approaches (see Fig.~\ref{snapshots}). Of course, due to the inherent stochastic nature of the interacting particle system, the snapshots do not match exactly for each realization. To reach a quantitative assessment on the validity
of the SPDE~(\ref{stoch_part_eq}) to describe the spatio-temporal properties of the system in the continuum limit, we have computed various correlation functions for the system's steady state.  
The attainment of the steady state is assessed by computing the long time evolution of the densities and various snapshots as those of
 Fig.~\ref{snapshots}. When the densities are found not to fluctuate significantly around  their average values and the
snapshots display statistically the same robust features at various times (typically $t\sim 100-1000$), the system is considered to be settled in its (reactive) steady state.

\begin{figure}
\begin{center} 
\includegraphics[scale=1]{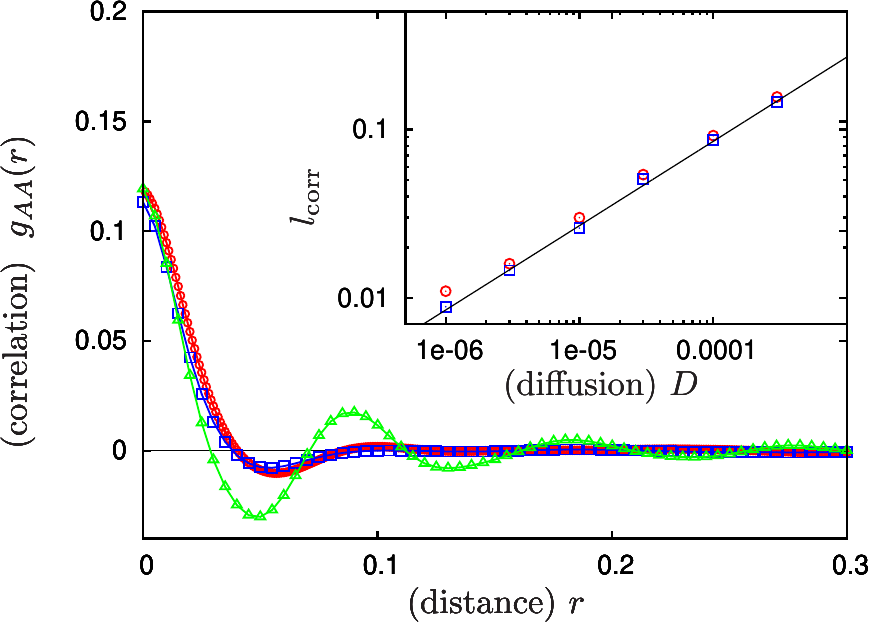}
\caption{\small Correlation functions. The spatial correlation $g_{AA}(r)$ as function of $ r$ in the reactive steady state is shown. 
We report results obtained  from stochastic simulations (red circles, for a lattice of linear size $L=1000$) and numerical solutions of the SPDE~(\ref{stoch_part_eq}), blue squares, and notice an excellent agreement. In both cases, results have been obtained for a value $D=3\times 10^{-6}$ and $\mu=\sigma=1$.  The typical correlation length $\ell_\text{corr}$ as a function of the diffusion constant $D$ is shown in the inset (on a double logarithmic scale). The scaling relation $\ell_\text{corr}\sim \sqrt{D}$, indicated by a black line, is clearly confirmed. We have also reported the results for the static correlation function $g_{AA}(r)$  of the patterns predicted by the 
deterministic PDE (green triangles); see text. The latter are found to be markedly less damped than those 
arising in the stochastic descriptions of the system.
\label{space_corr}} 
\end{center}                
\end{figure} 
We first consider equal-time correlation functions, which yield information about the size of the emerging spirals. As an example, we focus on the correlation $g_{AA}(|{\bm r-\bm r}'|)$ at ${\bm r}$ and ${\bm r'}$ of the subpopulation $A$, 
$g_{AA}(|{\bm r-\bm r}'|)=\langle a({\bm r},t)a({\bm r}',t) \rangle - \langle a({\bm r},t)\rangle \langle a({\bm r}',t) \rangle$. Here, the brackets $\langle  ...\rangle$ stand for an average over all histories.
In the steady state, the time dependence drops out and, because of translational and rotational invariance, the latter depends only on the separating distance $|{\bm r}-{\bm r'}|$. In Fig.~\ref{space_corr}, we report results  for $g_{AA}$ obtained from lattice simulations (red circles) and from
numerical solutions of the SPDE~(\ref{stoch_part_eq}) (blue squares), finding an excellent agreement between them. When the separating distance vanishes, the correlation reaches its maximal value and then decreases, exhibiting (damped) spatial oscillations. The latter reflect the underlying spiralling spatial structures, where the three  subpopulations alternate in turn. Damping results from the averaging over many small spirals. 
As described in the previous subsection, the correlation functions are characterized by their correlation length $\ell_\text{corr}$, which conveys information on the  typical size of the spirals.  In the inset of Fig.~\ref{space_corr}, we show the dependence of the correlation length on the diffusion rate $D$ in a double logarithmic plot which confirms  the scaling relation $\ell_\text{corr}\sim \sqrt{D}$, also inferred from general considerations.

\begin{figure}
\begin{center}  
\includegraphics[scale=1]{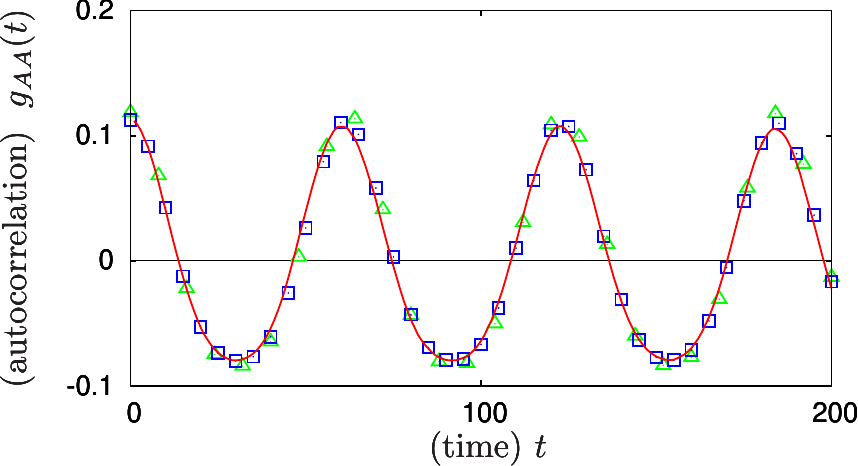}
\caption{\small Autocorrelation. We show the correlation $g_{AA}(t)$ as a function of time $t$. Results from stochastic simulations (red/gray line) are compared with those obtained from the numerical solutions of the SPDE (blue squares), as well as with those computed from the deterministic PDE (green triangles). All results are in excellent agreement with each other and are characterized by  oscillations at frequency $\Omega^\text{num}\approx 0.103$ (for $\mu=\sigma=1$); the latter is independent from the value of the diffusion $D$. These oscillations reflect the rotation of the spiral waves. The results from the SPDE and deterministic PDE have been obtained using $D=10^{-5}$, while stochastic simulations have been performed on a lattice of length $L=300$ with $D=10^{-4}$.
\label{time_corr}} 
\end{center}                
\end{figure}
We now consider the time dependence of the correlation functions and study the autocorrelation $g_{AA}(|t-t'|)$ of subpopulation $A$ at times $t$ and $t'$, for a fixed spatial position. This quantity is given by $g_{AA}(|t-t'|)=\langle a({\bm r},t)a({\bm r},t') \rangle - \langle a({\bm r},t)\rangle \langle a({\bm r},t') \rangle$
%
and only depends on the time difference $|t-t'|$. Both lattice simulations and SPDE~(\ref{stoch_part_eq}) yield oscillating correlation functions, as shown in Fig.~\ref{time_corr}. This periodic behavior, with a frequency numerically found to be $\Omega^\text{num}\approx 0.103$ (for $\sigma=\mu=1$), stems from the 
rotational nature of the spiral waves and is independent of the diffusion constant $D$. Below, this value is compared with an analytical prediction inferred from a deterministic description of the spatial system.
In the time intervals which we have investigated, $t\sim 1,000$, the oscillations, as reported in Fig.~\ref{time_corr}, are undamped. Therefore, on this time-scale, the position of the spirals' vortices  is stable in the steady state and not influenced by noise. On larger time-scales, however, we expect the vortices to perform random walks (see~\citep{cross-1993-65} for a general discussion as well as~\citep{tainaka-1994-50,nishiuchi-2008-387} for investigations of vortex dynamics in rock-paper-scissors models), with associated vortex annihilation and creation processes. Studies exploring such a behavior are promising for further broadening the understanding of stochastic effects on nonequilibrium steady state.  In the following, we consider another footprint of stochastic fluctuations: we show how the latter significantly influence the steady state of the system when starting from homogeneous initial densities.

\subsection{The role of stochasticity}

So far, we have considered stochastic descriptions, relying on lattice simulations and the stochastic partial differential equations~(\ref{stoch_part_eq}). In the latter, the noise terms are proportional  to $1/\sqrt{N}$, i.e. vanishing in the limit of $N\rightarrow\infty$. It is therefore legitimate to ask whether it is possible to simply neglect noise and describe the system
in terms of deterministic partial differential equations (PDE)~(\ref{det_part_eq}) given in Subsection~\ref{subsec:cont_limit}.
\begin{figure}
\begin{center}
\includegraphics[scale=1]{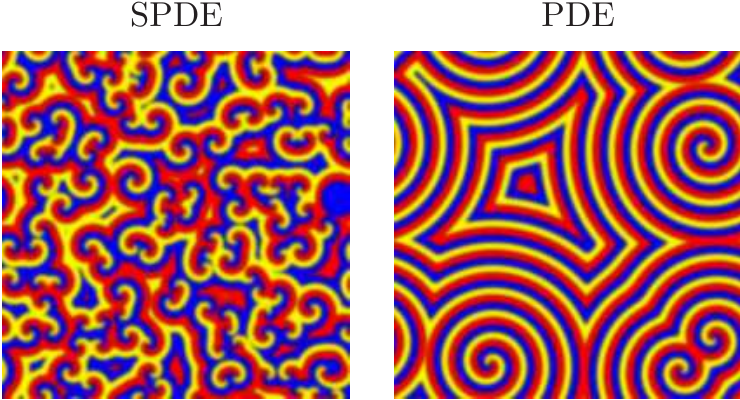}
\caption{\small The role of noise. Left: numerical solution of the SPDE~(\ref{stoch_part_eq}), starting from an homogeneous initial state (all subpopulations have an initial density $1/4$). Each color (level of gray) represents a different species (black dots denote empty spots). Right: if we ignore the noise terms of the SPDE~(\ref{stoch_part_eq}) and study the resulting deterministic PDE~(\ref{det_part_eq}), the steady state depends on the initial configuration. Here, this is illustrated by a snapshot of the steady state evolving  from the initial condition: $a(0)=a^*+\cos(2\pi xy)/100,~ b(0)=b^*,~c(0)=c^*$, see text. In both panels, the diffusion constant is $D=3\times 10^{-6}$ and $\sigma=\mu=1$.
\label{pict_stoch_det}} 
\end{center}                
\end{figure}  
To address this question and to reach a better understanding of the effects of  internal noise, we have numerically solved the deterministic PDE~(\ref{det_part_eq}) for various initial conditions. Of course, the latter have always to be spatially inhomogeneous, otherwise spatial patterns cannot emerge. Starting from a spatially inhomogeneous distribution of the populations, the deterministic equations are found to evolve towards a reactive steady state also characterized by the emergence of spirals. We have numerically
checked that these spirals' wavelengths $\lambda$ are the same as those obtained in the stochastic lattice simulations
and for the solutions of the SPDE~(\ref{stoch_part_eq}). 

However, there are major differences between the deterministic and the stochastic descriptions of the spatially extended system.  Concerning the SPDE, when the initial densities are the ones corresponding to the unstable internal fixed point of the rate equations~(\ref{ml_rate_eqs}), even without initial inhomogeneities, an entanglement of spiral waves arises in the course of time evolution only due to noise. We have numerically found that the latter is characterized by a universal vortex density of about $0.5$ per square wave length. For the deterministic PDE, spatially homogeneous initial conditions do not yield spiralling patterns. When starting with initial spatial inhomogeneities, the density of the latter sensitively determines the density of spirals, which can be much lower than in the stochastic situation. As an illustration, in Fig.~\ref{pict_stoch_det} we compare snapshots of the steady state of the SPDE and of the PDE. For the latter, we have chosen the initial condition $a(t=0)=a^*+\cos(2\pi xy)/100,~ b(0)=b^*,~c(0)=c^*$, just adding a small local perturbation to the value of the homogeneous fixed point $(a^*, b^*, c^*)$. While a large number of spirals cover the system in the stochastic case (Fig.~\ref{pict_stoch_det}, left), only four spirals appear in the deterministic situation; Fig.~\ref{pict_stoch_det} right.
This difference  is also manifest when one considers 
the spatial dependence of the correlation functions, as shown in Fig.~\ref{space_corr}. 
These quantities share the same maxima and minima for the stochastic and deterministic descriptions, which 
can be traced back to the fact that spirals have the same wavelengths, not affected by the noise.
However, the correlations obtained from lattice simulations and from the  SPDE are much more strongly damped than in the case of a deterministic PDE. This stems from  the fact that stochasticity effectively acts as an internal source of spatial inhomogeneities
(randomly distributed), resulting in a larger number of spirals (of small size). The agreement between the temporal dependence of the correlation functions in the deterministic and stochastic descriptions (see Fig.~\ref{time_corr}) is another consequence of the  fact that the spirals' frequency is not affected by the noise.

We can now ask what happens if one solves the SPDE~(\ref{stoch_part_eq}) with inhomogeneous initial conditions. To answer this question, we have systematically studied the emerging steady state upon varying the strength of noise and the degree of spatial inhomogeneity of the initial configuration. When these effects are of comparable intensity (i.e. when the noise strength is of the same order as the initial deviations of the densities from spatial homogeneity), the steady state is still driven by noise and gives rise to an entanglement of small spirals. In this situation,  the  universal density of $0.5$ spiral vortices per square wavelength arises. On the other hand, if the degree of inhomogeneity of the initial state is significantly higher than the noise level, the former dominates the systems' behavior
and the density of spirals is determined by the spatial structure of the initial configuration, as in Fig.~\ref{pict_stoch_det} (right). Therefore, if no initial inhomogeneities are present, stochasticity acts as a random source of inhomogeneities leading to the emergence of an entanglement of stable spiral vortices. The latter are stable against stochastic effects, as reflected by the undamped temporal oscillations of the autocorrelation function of  Fig.~\ref{time_corr}. On the other hand, if initial inhomogeneities exceed the noise level, they are responsible for the formation of vortices before noise can influence the system. 

From the above discussion, we infer that the range of initial conditions where noise influences the system's fate and leads to the universal density of about $0.5$ spiral vortices per square wavelength is rather small. It contains initial densities that lie around the values  of the unstable reactive fixed point of the rate equations~(\ref{ml_rate_eqs}), with spatial variations whose amplitude does not exceed the noise level. Other initial conditions, with more strongly pronounced spatial inhomogeneities, lead to spiral vortices that are determined by the initial inhomogeneities, and not stochastic effects. The latter behavior is remarkable, as it corresponds to a \emph{memory} of the system: The information about the initial state, i.e. its spatial inhomogeneities, is preserved in the reactive steady state where it manifests itself in the  position of the spirals' vortices. 
In the time interval that we have considered ($t\sim 1000$), the latter are stable during the temporal evolution and noise does not affect their location.
This feature is intimately connected to the nonequilibrium nature of the system. Indeed, in noisy equilibrium systems, the steady state does not convey information about the initial conditions. This stems from the standard assumption of ergodicity, as first formulated in the works of Boltzmann and Gibbs~\citep{Boltzmann,Gibbs}.  Only a genuine nonequilibrium steady state can display memory of the initial state.

Finally,  let us  comment on the nature of the noise encountered in our analysis. In the SPDE~(\ref{stoch_part_eq}), the noise is multiplicative because its strength depends on the densities of the subpopulations. Approximating  the latter by additive white noise [e.g., substituting $(a,b,c)$ by the values $(a^*,b^*,c^*)$ in  the expressions of the noise contributions], we essentially obtain the same results as with multiplicative noise. We understand this 
 behavior as stemming from the fact that  the main influence of noise is  to spatially and randomly perturb the system's initial state. Hence, as long its intensity scales like $N^{-1/2}$ (weak noise), the
fact that noise is either multiplicative or additive has no significant impact on the system's fate.

\subsection{The spirals' velocities, wavelengths, and frequencies}

Above, we have found that characteristic properties of the emerging spiral waves, like their wavelength and frequency, are unaffected by noise. To compute these quantities analytically, it is therefore not necessary to take noise into account, and we may focus on the study of the deterministic PDE~(\ref{det_part_eq}). In Subsection~\ref{subsec:CGLE}, we show how the dynamics of the latter is essentially captured by an appropriate complex Ginzburg-Landau equation (CGLE), given by Eq.~(\ref{CGLE_eq}) for the case under consideration here. The CGLE~(\ref{CGLE_eq}) allows to derive analytical results for the emergence  of spiral waves, their stability
and their  spreading velocity, as well as their wavelength and frequency.  We detail these findings in Subsections~\ref{subsec:velocity} and~\ref{subsec:wavelength}. Here, we  assess the accuracy and validity of these analytical predictions
 by comparing them  with values obtained from the numerical solutions of the SPDE~(\ref{stoch_part_eq}).

\begin{figure}
\begin{center}  
\includegraphics[scale=1]{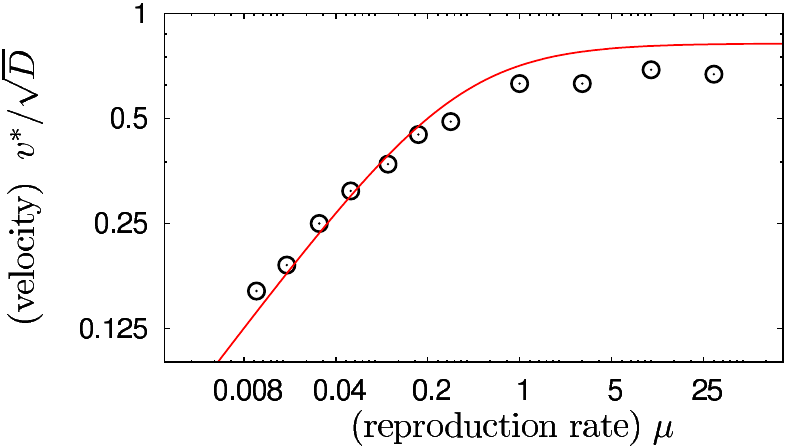}
\caption{\small Spreading velocity. We report the dependence of front velocity $v^*$ (rescaled by a factor $\sqrt{D}$) on the reproduction rate $\mu$. The time scale is set by keeping $\sigma=1$. In red (full line), we report the analytical predictions (\ref{vel}) obtained from the CGLE, which are compared
with numerical results (black dots). The latter are obtained from the numerical solutions of the SPDE~(\ref{stoch_part_eq}).\label{beta_velocity}}
\end{center}                
\end{figure} 
Let us first consider the spreading velocity $v^*$ of the emerging wave fronts. The analytical value, inferred from the CGLE~(\ref{CGLE_eq}) and derived in Subsection~\ref{subsec:velocity} [see Eq.~(\ref{vel})], reads $v^*=2\sqrt{c_1 D}$, where $c_1=\mu\sigma/[2(3\mu+\sigma)]$ is a coefficient appearing in the CGLE~(\ref{CGLE_eq}).
In numerical computations, the front velocity is obtained from the wavelength $\lambda$ and the frequency $\Omega$
of the emerging spirals. Namely, the wavelength $\lambda^\text{num}$ can be inferred from snapshots (as in Fig.~\ref{snapshots}), and the frequency $\Omega^\text{num}$ is computed from the oscillations of the autocorrelation (as in Fig.~\ref{time_corr}). The velocity then follows via $v^\text{num}=\lambda^\text{num}\Omega^\text{num}/2\pi$. As the wavelength is proportional to $\sqrt{D}$ and the frequency does not depend on the diffusion constant, one can easily check that the relation $v^\text{num}=\lambda^\text{num}\Omega^\text{num}/2\pi$ confirms that  $v^\text{num}\sim\sqrt{D}$, as in Eq.~(\ref{vel}).
In Fig.~\ref{beta_velocity}, we compare the analytical prediction (\ref{vel}) for $v^*$ with results obtained from the numerical solution of the SPDE~(\ref{stoch_part_eq}),
as function of the reproduction rate $\mu$ (setting $\sigma=1$, we fix the time scale), and find  a good agreement.
On the one hand, for small values of $\mu$ (much lower than the selection rate, $\mu\ll 1$), reproduction is the dominant limiter of the spatio-temporal evolution. In the limit $\mu\to 0$, the front velocity therefore only depends on $\mu$. From dimensional analysis, it follows $v^*\sim \sqrt{\mu}$, as  also confirmed by the analytical solution Eq.~(\ref{vel}). 
On the other hand, if reproduction is much faster than selection, $\mu\gg 1$, the latter limits the dynamics, and we recover $v^*\sim \sqrt{\sigma}$. In Fig.~\ref{beta_velocity}, as $\sigma=1$, this behavior translates into 
 $v^*$ being independent of $\mu$ in this limit.
While the numerical and analytical results coincide remarkably for low reproduction rates (i.e. $\mu\leq 0.3$), systematic deviations ($\approx 10\%$) appear at higher values. As an example, when selection and reproduction rates are equal, $\sigma=\mu=1$ (as was considered throughout the last section), we have numerically found a velocity $v^\text{num}\approx 0.63\sqrt{D}$, while   Eq.~(\ref{vel}) yields the analytical result $v^*=\sqrt{D/2}\approx 0.71\sqrt{D}$.

\begin{figure}
\begin{center}  
\includegraphics[scale=1]{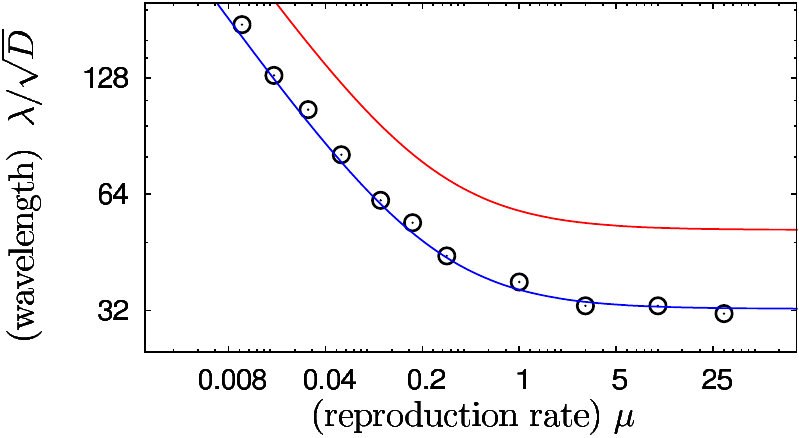}
\caption{\small The spirals' wavelength. We show the functional dependence of the wavelength $\lambda$ on the rate $\mu$ (with $\sigma=1$), and compare numerical results (black circles), obtained from the numerical solutions of the SPDE~(\ref{stoch_part_eq}), to analytical predictions (red line). The latter stem from the CGLE and are given by Eq.~(\ref{lambda}). They differ from the numerics by  a factor of $1.6$, see text. Adjusting this factor, c.f. the blue line,  the functional dependence is seen to agree very well with numerical results. \label{beta_lambda}}
\end{center}                
\end{figure} 
Concerning the spirals' wavelengths and frequencies, in Subsection~\ref{subsec:wavelength}, we analytically infer predictions from the CGLE~(\ref{CGLE_eq}) given by Eqs.~(\ref{omega}) and (\ref{lambda}).
We have checked these  results against numerical computations. 
In Fig.~\ref{beta_lambda}, the analytical estimates for the  wavelength $\lambda$ are compared with those obtained from the  numerical solution of the SPDE~(\ref{stoch_part_eq}) for different values of the reproduction rate $\mu$. We notice that there is an excellent agreement between analytical and numerical results for the functional dependence of $\lambda$ on $\mu$. 
 For low reproduction rates ($\mu\ll 1$) we have $\lambda\sim 1/\sqrt{\mu}$, while
when reproduction occurs much faster than selection ($\mu\gg 1$), the dynamics is independent of $\mu$ and $\lambda\sim 1/\sqrt{\sigma}$.
We have also found that
the analytical result predicts an amplitude of $\lambda$ which exceeds that obtained from numerical computations by a constant factor $\approx 1.6$, taken into account in Fig.~\ref{beta_lambda}. We attribute this deviation to the fact  that the CGLE~(\ref{CGLE_eq}) (stemming from the normal form~(\ref{normal_form})) describes a dynamics exhibiting a limit cycle, while the full May-Leonard rate equations~(\ref{ml_rate_eqs}) are characterized by heteroclinic orbits. The correct functional dependence of the wavelength $\lambda$ on the reproduction rate $\mu$  is therefore especially remarkable.
Elsewhere it  will be shown that in the presence of mutations, inducing a limit-cycle behavior, the description of the emerging spiral waves in terms of CGLE~(\ref{CGLE_eq}) becomes fully accurate.

For the spirals' frequency, we analytically obtain $\Omega= \Omega(q^\text{sel})=\omega+2(c_1/c_3)\left(1-\sqrt{1+c_3^2}\right)$, see Subsection~\ref{subsec:wavelength}. As already inferred from numerical simulations (Sec. III.B), $\Omega$ does not depend on the diffusion $D$. Quantitatively, and as an example for $\mu=\sigma =1$, we obtain the analytical prediction $\Omega\approx 0.14$, which differs by a factor $\approx 1.4$ from the numerical value $\Omega^\text{num}\approx 0.103$ found in Fig.~\ref{time_corr}. As for the wavelength, this difference stems from the fact that the May-Leonard rate equations~(\ref{ml_rate_eqs}) predict heteroclinic orbits approaching the boundaries of the phase space, while the dynamics underlying the CGLE is characterized by limit cycles (usually distant from the edges of the phase space) resulting from a (supercritical) Hopf bifurcation.

\section{Theory \label{sec:theory} }

In the following, we present and discuss an analytical approach built on stochastic partial differential equations (SPDE) arisen in the proper continuum limit of large systems with mobile individuals. Applying the theory of invariant manifolds and normal forms to the nonlinear parts of these equations, we show that the SPDE fall into the universality class of a complex Ginzburg-Landau equation (CGLE). We derive the latter and employ it for a  quantitative determination of properties of the spiral waves.

\subsection{Rate equations}
\label{subsec:RE}

The deterministic rate equations (RE) describe the temporal evolution of the stochastic lattice system, defined by the reactions~(\ref{ml_react}), in a mean-field manner, i.e. they neglect all spatial correlations. They may be seen as a deterministic description (for example emerging in the limit of large system sizes) of systems without spatial structure, such as Moran processes \citep{moran-1958-54,traulsen-2005-95,traulsen-2006-74} or urn models \citep{Feller,reichenbach-2006-74}. The study of the RE is the ground on which the analysis of the stochastic spatial system is built. In particular,  the properties of the RE are  extremely useful for the derivation of the system's SPDE~(\ref{stoch_part_eq}) and CGLE~(\ref{CGLE_eq}).
 
Let $a,~b,~c$ denote the densities of subpopulations $A,~B$, and $C$, respectively. The overall density $\rho$ then reads $\rho=a+b+c$. As every lattice site is at most occupied by one individual, the overall density (as well as  densities of each subpopulation) varies between $0$ and $1$, i.e. $0\leq \rho\leq 1$.
With these notations, the RE for the reaction~(\ref{ml_react}) are given by
\begin{eqnarray}
\partial_t a&=&a[\mu(1-\rho)-\sigma c]\,,\cr
\partial_t b&=&b[\mu(1-\rho)-\sigma a]\,,\cr
\partial_t c&=&c[\mu(1-\rho)-\sigma b]\,.
\label{ml_rate_eqs}
\end{eqnarray}
They imply the following temporal evolution of the total density:
\begin{equation}
\partial_t \rho = \mu\rho(1-\rho)-\sigma(ab+bc+ac)\,.
\label{ml_dens}
\end{equation}
These equations have been introduced and investigated by~\citet{may-1975-29}. In the following, we review some of their properties.

Equations~(\ref{ml_rate_eqs}) possess four  absorbing fixed points. One of these (unstable) is  associated with the extinction of all subpopulations, $(a^*_1,b^*_1,c^*_1)=(0,0,0)$. The others are heteroclinic points (i.e. saddle points underlying the heteroclinic orbits) and correspond to the survival of only one subpopulation, $(a^*_2,b^*_2,c^*_2)=(1,0,0),(a^*_3,b^*_3,c^*_3)=(0,1,0)$ and $(a^*_4,b^*_4,c^*_4)=(1,0,0)$, shown in blue (dark gray) in Fig.~\ref{pict_coord}.  In addition, there exists a fixed point, indicated in red (gray) in Fig.~\ref{pict_coord}, where all three subpopulations coexist (at equal densities), namely $(a^*,b^*,c^*)=\frac{\mu}{3\mu+\sigma}(1,1,1)$. 
For a non-vanishing selection rate, $\sigma>0$, \citet{may-1975-29} showed that the reactive fixed point is unstable, and the system  asymptotically approaches the boundary of the phase space (given by the  planes $a=0,\, b=0$, and $c=0$). There, they observed \emph{heteroclinic orbits}: the system oscillates between states  where nearly only one subpopulation is present, with rapidly increasing cycle duration. While mathematically fascinating, this behavior was recognized to be unrealistic~\citep{may-1975-29}. For instance, in a biological setting,  the system will, due to finite-size fluctuations, always reach one of the absorbing fixed points in the vicinity of the heteroclinic orbit, and then only one population survives.

Linearization of the RE~(\ref{ml_rate_eqs}) around the reactive fixed point leads to $\partial_t
{\bm x}=\mathcal{A}{\bm x}$ with the vector ${\bm x}=(a-a^*,b-b^*,c-c^*)^T$ and the Jacobian matrix
\begin{equation} 
 \mathcal{A}=-\frac{\mu}{3\mu+\sigma}\begin{pmatrix} \mu & \mu & \mu+\sigma \\ \mu+\sigma & \mu & \mu \\  \mu & \mu+\sigma & \mu \end{pmatrix} \,.
\end{equation}
As this matrix is circulant, its eigenvalues can be obtained from a particularly simple general formula (see e.g. \citep{Hofbauer}); they
read:
\begin{eqnarray} 
\lambda_0&=&-\mu\,, \cr
\lambda_1&=&\frac{1}{2}\frac{\mu\sigma}{3\mu+\sigma}\big[1+\sqrt{3}i\big]\,, \cr
\lambda_2&=&\frac{1}{2}\frac{\mu\sigma}{3\mu+\sigma}\big[1-\sqrt{3}i\big]\,.
\end{eqnarray}
This shows that the  reactive fixed point is stable along the eigendirection of the first eigenvalue $\lambda_0$. As elaborated below, there exists an invariant manifold~\citep{Wiggins} (including the reactive fixed point),  that the system quickly approaches. To first order such a manifold is the plane normal to the eigendirection of $\lambda_0$. On this invariant manifold, flows spiral away from the reactive fixed point, which is an unstable focus, as sketched in Fig.~\ref{pict_coord} (blue trajectory). 

The linear stability analysis only reveals  the local stability of the fixed points. 
The global instability of the reactive fixed point is proven by the existence of a Lyapunov function $\mathcal{L}$ \citep{Hofbauer,may-1975-29}:
\begin{equation}
\mathcal{L}=\frac{abc}{\rho^3}\,.
\end{equation}
In fact, using Eqs.~(\ref{ml_rate_eqs}) and~(\ref{ml_dens}), the time derivative of $\mathcal{L}$ is found to be always non-positive,
\begin{equation}
\partial_t\mathcal{L} =-\frac{1}{2}\sigma \rho^{-4}abc\big[(a-b)^2+(b-c)^2+(c-a)^2\big] \leq 0~.
\label{lyapunov_time}
\end{equation} 
We note that $\partial_t\mathcal{L}$ vanishes only at the boundaries ($a=0,b=0$ or $c=0$) and along the line of equal densities, $a=b=c$. The latter coincides with the eigendirection of $\lambda_0$, along which the system approaches the reactive fixed point. However, on the invariant manifold we recover $\partial_t\mathcal{L}<0$, corresponding to a globally unstable reactive fixed point, as exemplified by the trajectory shown in Fig.~\ref{pict_coord}.

As the RE (\ref{ml_rate_eqs}) have one real eigenvalue smaller than zero and a pair of complex conjugate eigenvalues, they fall into the class of the Poincar$\acute{\text{e}}$-Andronov-Hopf bifurcation, well known in the mathematical literature \citep{Wiggins}.
The theory of invariant and center manifolds allows us to recast these equations into a normal form. The latter will turn out to be extremely useful in the derivation of the CGLE. In the following, we  derive the invariant manifold to second order  as well as the normal form of the RE.

\subsection{Invariant manifold}
\label{subsec:IM}

An invariant manifold is a subspace, embedded in the phase space, which is left invariant by the RE (\ref{ml_rate_eqs}), i.e.
by the deterministic dynamics. In the phase space, this means that flows starting on an invariant manifold always lie and evolve on it.  
Here, we consider a two-dimensional invariant manifold associated with the reactive fixed point of the RE~(\ref{ml_rate_eqs}) onto which all trajectories (initially away from the invariant manifold) decay exponentially quickly (see below). Upon restricting the dynamics to that appropriate invariant manifold, the system's degrees of freedom are reduced from three to two, which greatly simplifies the mathematical analysis.

To determine this invariant manifold, we notice that the eigenvector of the eigenvalue  $\lambda_0<0$ at the reactive fixed point is a stable (attractive) direction. Therefore, to lowest order around the reactive fixed point,
the invariant manifold is simply the plane normal to the eigendirection of $\lambda_0$.
As well known in the theory of dynamical systems, beyond first order, nonlinearities affect the invariant manifold, which is therefore only tangent to this plane. To include nonlinearities,  it is useful to  introduce suitable coordinates $(y_A,y_B,y_C)$ originating in 
 the reactive fixed point. We choose the $y_C$-axis to coincide with the eigenvector of $\lambda_0$,
such that  the plane $y_C=0$ is the invariant manifold to linear order, onto which the flows of (\ref{ml_rate_eqs}) relax quickly. 
The coordinates $y_A$ and $y_B$ are chosen to span the plane normal to the axis $y_C$,
 forming an orthogonal set. Such  coordinates ${\bm y}=(y_A,y_B,y_C)$ are, e.g.,  obtained by the linear transformation $\bm{ y}=\mathcal{S}\bm{ x}$, with the matrix $\mathcal{S}$ given by
\begin{equation}
\mathcal{S}=\frac{1}{3}\begin{pmatrix} \sqrt{3} & 0 & -\sqrt{3} \\
                                        -1 & 2 & -1 \\
					1 & 1 & 1          \end{pmatrix} \,,
\end{equation}
where $\bm{x}=(a-a^*,b-b^*,c-c^*)^T$ denotes the deviation of the densities from the reactive fixed point.
The coordinates $(y_A,y_B,y_C)$ are shown in Fig.~\ref{pict_coord} and  their  equations 
of motion are given  in Appendix A.

The May-Leonard model is symmetric under cyclic permutations of $A,B$, and $C$.
In the $\bm{ y}$-coordinates,  each of these permutations translates into a rotation 
of $2\pi/3$ around the $y_C$ axis. The equations of motion reflect this symmetry of the system, as can be checked explicitly in the Eqs.~(\ref{eq_yaybyc}) in Appendix A. 

To parameterize the  invariant manifold sketched in Fig.~\ref{pict_coord}, we seek a function $G(y_A,y_B)$, with $y_C=G(y_A,y_B)$. If all nonlinearities of the RE are taken into account, this is a very complicated problem. However, for our purpose it is sufficient to  expand $G$   to  second order in $y_A,y_B$. As the invariant manifold is left invariant by the RE, by definition, $G$ must obey
\begin{equation}
\partial_tG\big(y_A(t),y_B(t)\big)=\frac{\partial G}{\partial y_A}\partial_t y_A + \frac{\partial G}{\partial y_B}\partial_t y_B = \partial_t y_C\Big|_{y_C=G}~.
\label{eq_G}
\end{equation}
To  linear order in $y_A$ and $y_B$, we simply have $G=0$ and recover $y_C=0$, corresponding to  the plane normal to the $y_C$-direction. We have anticipated this result above: to first order, the invariant manifold coincides with this plane, and is tangential to it when higher orders are included. To second order, only linear terms of $\partial_t y_A,\partial_ty_B$ contribute to Eq.~(\ref{eq_G}). The latter are invariant under rotations in the $(y_A,y_B)$-plane, and $G$ must obey the same symmetry. It is therefore proportional to $y_A^2+y_B^2$. 
After some calculations, detailed in Appendix B,  one obtains:
\begin{equation}
y_C=G(y_A,y_B)=\frac{\sigma}{4\mu}\frac{3\mu+\sigma}{3\mu+2\sigma}(y_A^2+y_B^2) + o({\bm y}^2)\,.
\label{center_eq}
\end{equation} 
The comparison of this expression for the invariant manifold, valid to second order, with the numerical solutions 
of the RE~(\ref{ml_rate_eqs}) (which should, up to an initial transient, lie on the invariant manifold) confirms that (\ref{center_eq}) is an accurate approximation, with
only minor deviations occurring near the boundaries 
of the phase space.

\subsection{Normal form}
\label{subsec:NF}

Nonlinear systems are notably characterized by the  bifurcations that they exhibit~\citep{Wiggins}.
Normal forms are defined as the simplest differential equations that capture the essential features of a system near a bifurcation point, and therefore provide insight into the system's universal behavior. 
Here, we derive the normal form associated with the RE~(\ref{ml_rate_eqs}) of the May-Leonard model and show
that they belong to the universality class of the Hopf bifurcation~\citep{Wiggins}. Below, we demonstrate that this property allows
to describe the system in terms of a well-defined complex Ginzburg-Landau equation.

Restricting the (deterministic) dynamics onto the invariant manifold, given by Eq.~(\ref{center_eq}), the system's behavior can be
analyzed  in terms of two variables. Here, we choose to express $y_C$ as a function of $y_A$ and $y_B$, with the resulting rate equations (up to  cubic oder) given in Appendix A.
The latter can be cast into a normal form (see \citep{Wiggins} Chapter 2.2) by performing a nonlinear variable transformation $\bm{ y}\rightarrow \bm{ z}$ which eliminates the quadratic terms and preserves the linear ones (i.e.  $\bm{y}$ and $\bm{z}$ coincide to linear order). As an ansatz for such a transformation, we choose the most general  quadratic expression in $\bm{y}$ for the new variable $\bm{z} $. Details of the calculation can be found in Appendix C. Here we quote the result for the normal form of the RE in the new variables:
\begin{align}
\partial_t z_A =&~c_1 z_A + \omega z_B - c_2\big(z_A+c_3z_B\big)(z_A^2+z_B^2) + o({\bm z}^3)\,,\cr
\partial_t z_B =&~c_1 z_B - \omega z_A - c_2\big(z_B-c_3z_A\big)(z_A^2+z_B^2) + o({\bm z}^3)\,.
\label{normal_form}
\end{align}
In these equations,  
\begin{equation}
\omega=\frac{\sqrt{3}}{2}\frac{\mu\sigma}{3\mu+\sigma}\,,
\end{equation}
is the (linear) frequency of oscillations around the reactive fixed point. The constant
\begin{equation}
c_1=\frac{1}{2}\frac{\mu\sigma}{3\mu+\sigma}\,,
\label{c1}
\end{equation}
gives the intensity of the linear drift away from the fixed point, while 
\begin{align}
c_2&=\frac{\sigma(3\mu+\sigma)(48\mu+11\sigma)}{56\mu(3\mu+2\sigma)}\,,
\label{c2}\\
c_3&=\frac{\sqrt{3}(18\mu+5\sigma)}{48\mu+11\sigma}\,,
\label{c3}
\end{align}
are the coefficients of the cubic corrections.

To gain some insight into the dynamics in the normal form, it is useful to rewrite (\ref{normal_form}) in
 polar coordinates $(r,\phi)$, where $z_A=r\cos\phi, z_B=r\sin\phi$. This leads to 
\begin{align}
\partial_tr~&=r [c_1-c_2r^2]~,\cr
\partial_t\theta~&=-\omega+c_2c_3r^2~.
\label{normal_polar}
\end{align}
These equations only have a radial dependence, which clearly reveals a polar symmetry.
They predict the emergence of a limit cycle of radius $r=\sqrt{c_1/c_2}$  and therefore fall into the universality class of the (supercritical) Hopf bifurcation.
However, when all nonlinearities are taken into account, the  RE~(\ref{ml_rate_eqs}) give rise to heteroclinic orbits instead of  limit cycles. The latter rapidly approach the boundaries of the phase space, and thus are in general well separated from the
limit cycles predicted by (\ref{normal_polar}). When comparing results inferred from the CGLE and stochastic lattice simulations in the results section, we have shown how  this causes some quantitative mismatch, stemming from the differences between
the solutions of (\ref{ml_rate_eqs}) and (\ref{normal_polar}). However,  we have also seen  that most features of the system are actually aptly captured by the normal form (\ref{normal_form}). Elsewhere, it will be shown that mutations
between subpopulations lead to limit cycles resulting from a Hopf bifurcation.

\subsection{Spatial structure and the continuum limit}
\label{subsec:cont_limit}

The system under consideration possesses spatial degrees of freedom, which are neither taken into  account in the RE~(\ref{ml_rate_eqs}) nor in the normal form ~(\ref{normal_form}). Here, within a proper continuum limit, we show how the spatial arrangement of mobile individuals may be included into the analytical description   by supplementing  the RE~(\ref{ml_rate_eqs})  with spatially dependent densities and diffusive terms~(\ref{det_part_eq}). When one additionally accounts for stochastic effects (internal noise, see below), the system is aptly described by a set of stochastic partial differential equations (SPDE) (\ref{stoch_part_eq}).

The reactions (\ref{ml_react}) and the exchange processes take place on 
a $d$-dimensional hypercubic lattice (with periodic boundary conditions) of linear size $L$ and 
 comprising $N=L^d$  sites. The coordination number  $z=2d$ gives
the number of nearest neighbors of each lattice site. We set the length of the lattice to unity, such that the distance between two nearest neighbors is  $\delta r=N^{-1/d}$. The density 
of subpopulations $A,~B$ and  $C$ at time $t$ and site  ${\bm r}=(r_1,...,r_d)$ is denoted
$a({\bm r},t),~b({\bm r},t)$, and $c({\bm r},t)$, respectively. According to the ``bimolecular'' reactions (\ref{ml_react}), the equations of motion of these quantities only involve neighbors located at  ${\bm r}\pm \delta r \cdot {\bm e}_i$, where  $\{{\bm e}_i\}_{i=1...d}$ is the basis of the lattice. Here, we first ignore all forms of correlation and fluctuations 
of the local density. While noise will be taken into account below, due to diffusion, correlations between neighboring sites vanish in the  continuum limit of large systems (see below).
We obtain the following equation for
the time evolution of its mean (average) value  $a({\bm r},t)$:
\begin{align}
&\partial_t a({\bm r},t)
=\frac{1}{z}\sum_{\pm,i=1}^d\Big\{2\epsilon\big[ a({\bm r}\pm \delta r \cdot {\bm e}_i,t)-a({\bm r},t)\big]\cr
&\quad+\mu a({\bm r}\pm \delta r \cdot {\bm e}_i,t)\big[1-a({\bm r},t)-b({\bm r},t)-c({\bm r},t)\big]\cr
&\quad-\sigma c({\bm r}\pm \delta r \cdot {\bm e}_i,t) a({\bm r},t)\Big\}\,.
\label{cont_a_time}
\end{align}
For an analytical description of the lattice model, fruitful insights are gained by considering a continuum limit with $N\rightarrow\infty$.
As the lattice size is kept fixed to $1$, in this limit the distance $\delta r$ between two neighboring sites 
approaches zero, i.e. $\delta r=N^{-1/d}\rightarrow 0$. This allows to treat ${\bm r}$ as a continuous variable
and the following expansion is justified:
\begin{align}
a({\bm r}\pm \delta r \cdot {\bm e}_i,t)=\,&a({\bm r},t)\pm \delta r \partial_i a({\bm r},t) \cr
&+ \frac{1}{2}\delta r^2 \partial_i^2 a({\bm r},t) + o(\delta r^2)\,.
\label{expansion_a}
\end{align}
With this expression, the first term on the right-hand-side (RHS) of Eq.~(\ref{cont_a_time}) becomes (up to second order)
\begin{align}
(2\epsilon/z)\sum_{\pm}\big[ a({\bm r}\pm \delta r \cdot {\bm e}_i,t)-a({\bm r},t)\big] = (\epsilon/d)\delta r^2\partial_i^2 a({\bm r},t).
\nonumber
\end{align}
If we rescale the exchange rate $\epsilon$ with the system size $N$ according to 
\begin{equation}
\epsilon=DdN^{2/d}\,,
\end{equation}
with a fixed (diffusion) constant $D$, we note that $\epsilon \delta r^2= Dd$. For the other terms on RHS of Eq.~(\ref{cont_a_time}), only the zeroth-order contributions in the expansion of $a({\bm r}\pm \delta r \cdot {\bm e}_i,t)$ do not vanish
when  $N\rightarrow\infty$ (i.e. $\delta r\rightarrow 0$).
In this continuum limit, Eq.~(\ref{cont_a_time}) thus turns into
\begin{align}
\partial_t a({\bm r},t)=&D\Delta a({\bm r},t)
+\mu  a({\bm r},t)\big[1- \rho({\bm r},t) \big]-\sigma a({\bm r},t)c({\bm r},t)\,,
\end{align}
wit the local density $\rho({\bm r},t)=a({\bm r},t)+b({\bm r},t)+c({\bm r},t)$.
The equations of motion for $b({\bm r},t)$ and $c({\bm r},t)$ are obtained similarly. 
We therefore obtain  the following set of partial differential equations (PDE):
\begin{align}
\partial_t a({\bm r},t)&= D\Delta a({\bm r},t)+\mu  a({\bm r},t)[1-\rho({\bm r},t)] -\sigma a({\bm r},t)c({\bm r},t)\,,\cr
\partial_t b({\bm r},t)&= D\Delta b({\bm r},t)+\mu  b({\bm r},t)[1-\rho({\bm r},t)] -\sigma b({\bm r},t)a({\bm r},t)\,,\cr
\partial_t c({\bm r},t)&= D\Delta c({\bm r},t)+\mu  c({\bm r},t)[1-\rho({\bm r},t)] -\sigma b({\bm r},t)c({\bm r},t) \,.
\label{det_part_eq}
\end{align}
The difference to the rate equations~(\ref{ml_rate_eqs}) lies in the spatial dependence through diffusive terms, proportional to the diffusion constant $D$.

\subsection{Noise}
\label{subsec:noise}

The discrete character  of the individuals involved in the May-Leonard reactions~(\ref{ml_react}) and the exchange processes are responsible for intrinsic stochasticity arising in the system. For the  treatment of this internal noise, we note a time-scale separation between the reactions~(\ref{ml_react})  and the exchange events. Namely, in the continuum limit  $N\rightarrow\infty$, according to (\ref{eps_scaling}) one has $\epsilon\rightarrow\infty$. This means that exchanges occur on a much faster time-scale than the reactions~(\ref{ml_react}). 
Consequently, a large number of exchange events occurs between two reactions and can be treated  deterministically. As shown below, the fluctuations associated with the exchange processes vanish as $1/N$ (for $N\rightarrow\infty$), while those stemming from (\ref{ml_react}) scale as $1/\sqrt{N}$. The latter are therefore the dominating source of noise, while the former can be neglected in the continuum limit. 
To establish these results, we consider large system sizes $N$  where a stochastic
description in terms of Fokker-Planck equations is generally appropriate~\citep{VanKampen,Gardiner}. The latter can be obtained from  Kramers-Moyal expansion (i.e. a system-size expansion) of the underlying master equation  \citep{Tauber}, see Appendix D. In Fokker-Planck equations, fluctuations are encoded in a noise matrix denoted $\mathcal{B}$. Equivalently, a set of  Ito stochastic (partial) differential equations (often referred to as Langevin equations) can be 
systematically derived. For these SPDE, the noise, often white, is encoded in the
``square root'' of the matrix $\mathcal{B}$. Namely, in this framework, the strength of fluctuations and the correlations are given by a matrix $\mathcal{C}$, defined as $\mathcal{C}\mathcal{C}^T=\mathcal{B}$. 
Below, we derive the relevant contributions to the noise matrices $\mathcal{B}$ and $\mathcal{C}$, which lead to the appropriate  stochastic partial differential equations (SPDE) of the system.

Following \citep{Gardiner} (Chapter 8.5), we first show that fluctuations stemming from pair exchanges scale as $1/N$. Consider two nearest neighboring lattice sites ${\bm r}$ and ${\bm r'}$. The rate for an  individual $A$ to hop from ${\bm r}$ to ${\bm r'}$ is given by $\epsilon z^{-1}a({\bm r})[1-a({\bm r'})]$ (for simplicity, we drop the time-dependence). Together with the reverse process, i.e. hopping from site ${\bm r'}$ to ${\bm r}$, this yields the non-diagonal part of $\mathcal{B}({\bm r},{\bm r'})$ (see e.g. \citep{Tauber}):
\begin{equation} 
\mathcal{B}({\bm r},{\bm r'}\neq {\bm r})=-\frac{\epsilon}{Nz}\big\{a({\bm r})[1-a({\bm r'})]+a({\bm r'})[1-a({\bm r})]\big\}\,.
\end{equation}
Similarly, the diagonal entries of $\mathcal{B}$ read
\begin{equation}
\mathcal{B}({\bm r},{\bm r})=\frac{\epsilon}{Nz}\sum_{\text{n.n.}\bm r''}\big\{a({\bm r})[1-a({\bm r''})]+a({\bm r''})[1-a({\bm r})]\big\}\,,
\end{equation}
where the sum runs over all nearest neighbors (n.n.) ${\bm r''}$ of the site ${\bm r}$.
It follows from these expressions that 
\begin{align}
\mathcal{B}&({\bm r},{\bm r'})=\frac{\epsilon}{Nz}\sum_{\text{n.n.}\bm r''}(\delta_{{\bm r},{\bm r'}}-\delta_{{\bm r'},{\bm r''}})\cr
&\times\big\{ a({\bm r})[1-a({\bm r''})]  +a({\bm r''})[1-a({\bm r})]\big\}\,.
\end{align}
In the continuum limit, with $\delta r\rightarrow 0$, we use the fact that $\delta_{{\bm r},{\bm r'}}\rightarrow \delta r^d\delta({\bm r}-{\bm r'})$ and obtain
\begin{align}
\mathcal{B}&({\bm r},{\bm r'})=\frac{\epsilon}{Nz}\delta r^d\sum_{\pm,i=1}^{d}\big[\delta({\bm r}-{\bm r}')-\delta({\bm r}\pm\delta r{\bm e}_i-{\bm r}')\big]\cr
&\times\big\{ a({\bm r})[1-a({\bm r}\pm\delta r{\bm e}_i)]+a({\bm r}\pm\delta r{\bm e}_i)[1-a({\bm r})] \big\}\,.
\end{align}
As in Eq.~(\ref{expansion_a}), we expand $\delta({\bm r}\pm\delta r{\bm e}_i-{\bm r}')$ and $a({\bm r}\pm\delta r{\bm e}_i)$ to second order and observe that only quadratic terms in $\delta r$ do not cancel.
With $\epsilon=DdN^{2/d}$ and $\delta r=N^{-1/d}$, we thus find:
\begin{align}
\mathcal{B}({\bm r},{\bm r'})=\frac{D}{N^2}\partial_{\bm r}\partial_{\bm r'}\big[\delta({\bm r}-{\bm r}')a({\bm r})(1-a({\bm r}) \big].
\end{align}
The noise matrix $\mathcal{B}$ of the Fokker-Planck equation associated with the exchange processes therefore scales
as $N^{-2}$. In the corresponding SPDE, the contribution to noise of the exchange processes scales like $N^{-1}$.

We now consider the fluctuations stemming from  May-Leonard reactions~(\ref{ml_react}). A detailed discussion of the treatment of fluctuations arising from discrete reactions in lattice systems can be found in Chapter 8 of Ref.~\citep{Gardiner}. Following the derivation therein, one recovers noise terms of the same form as in the corresponding non-spatial model, although the densities now have spatial dependence. Thus, they may be found via a straightforward Kramers-Moyal expansion of the master equation describing the well-mixed system, see e.g. \citep{traulsen-2005-95,traulsen-2006-74, reichenbach-2006-74}. 
Here, we report the results and relegate details of the derivation to Appendix D. As the reactions~(\ref{ml_react}) decouple birth from death processes, the noise matrices $\mathcal{B}$ and $\mathcal{C}$ are diagonal. In particular, the diagonal parts of $\mathcal{C}$ read
 \begin{align}   
\mathcal{C}_{A}&=\frac{1}{\sqrt{N}}\sqrt{a({\bm r},t)\big[\mu(1- \rho({\bm r},t))+\sigma c({\bm r},t)\big]}\,,\cr
\mathcal{C}_{B}&=\frac{1}{\sqrt{N}}\sqrt{b({\bm r},t)\big[\mu(1- \rho({\bm r},t))+\sigma a({\bm r},t)\big]}\,,\cr
\mathcal{C}_{C}&=\frac{1}{\sqrt{N}}\sqrt{c({\bm r},t)\big[\mu(1- \rho({\bm r},t))+\sigma b({\bm r},t)\big]}\,.
\label{C}  
\end{align} 
The SPDE for the densities $a({\bm r},t),b({\bm r},t),c({\bm r},t)$ are thus given by the partial differential equations~(\ref{det_part_eq}) supplemented by  the corresponding noise terms, which leads to:
 \begin{align}
\partial_t a({\bm r},t)= D\Delta a({\bm r},t)&+\mu  a({\bm r},t)[1-\rho({\bm r},t)]\cr
& -\sigma a({\bm r},t)c({\bm r},t) + \mathcal{C}_{A}\xi_A\,,\cr
\partial_t b({\bm r},t)= D\Delta b({\bm r},t)&+\mu  b({\bm r},t)[1-\rho({\bm r},t)]\cr
& -\sigma b({\bm r},t)a({\bm r},t) + \mathcal{C}_{B}\xi_B\,,\cr
\partial_t c({\bm r},t)= D\Delta c({\bm r},t)&+\mu  c({\bm r},t)[1-\rho({\bm r},t)]\cr
& -\sigma b({\bm r},t)c({\bm r},t) + \mathcal{C}_{C}\xi_C \,,
\label{stoch_part_eq}
\end{align}
where $\Delta$ denotes the Laplacian operator, and the Gaussian white noise terms $\xi_i({\bm r},t)$ have a spatio-temporal dependence, with the correlations
\begin{equation}
\langle \xi_i({\bm r},t)\xi_j({\bm r}',t')\rangle=\delta_{ij}\delta({\bm r}-{\bm r}')\delta(t-t')\,.
\label{SPDE}
\end{equation}

\subsection{Complex Ginzburg-Landau Equation (CGLE) }
\label{subsec:CGLE}

The reaction terms appearing in the SPDE~(\ref{stoch_part_eq}) coincide with those of the rate equations~(\ref{ml_rate_eqs}). Above, we have recast the latter in the  normal form~(\ref{normal_form}).   Applying the same transformations to the SPDE~(\ref{stoch_part_eq}) yields reaction terms as in (\ref{normal_form}). However, owed to the nonlinearity of the transformation, additional  nonlinear diffusive terms appear in the spatially-extended version of (\ref{normal_form}). In the following, the latter will be ignored.
 Furthermore,  when discussing the spatio-temporal properties of the system, we have encountered rotating spiral waves, whose velocity, wavelength and frequency have turned out to be unaffected by noise (in the continuum limit). 
An important consequence of this finding is that it is not necessary to take noise into account
to compute such quantities. This greatly simplifies the problem and, omitting any noise contributions, we  focus on 
two coupled partial differential equations which are conveniently rewritten
as a complex PDE in terms of the complex variable $z=z_A+iz_B$ (see Appendix C):
\begin{align}
\partial_t z({\bm r},t)=\,&D\Delta z({\bm r},t)+(c_1-i\omega)z({\bm r},t) \cr
&\quad -c_2(1-ic_3)|z({\bm r},t)|^2z({\bm r},t)\,.
\label{CGLE_eq}
\end{align}
Here, we recognize the celebrated  complex Ginzburg-Landau equation (CGLE), whose properties have been extensively studied~\citep{cross-1993-65,aranson-2002-74}. In particular, it is known that in two dimensions the latter gives rise to a broad range of coherent structures, including spiral waves whose  velocity, wavelength and frequency can be computed analytically. 

For the system under consideration, we can check that the CGLE~(\ref{CGLE_eq}) predicts the emergence of spiral waves which are stable against frequency modulation: no Benjamin-Feir or Eckhaus instabilities occur. 
As a consequence, one expects no intermittencies or spatio-temporal chaos, but only rotating spirals~\citep{aranson-2002-74}. 
In our discussion, we have verified the validity of this prediction for various sets of the parameters,
 both for stochastic lattice simulations and for the solutions of the SPDE (\ref{stoch_part_eq}).
In particular, the expression (\ref{CGLE_eq}) with the parameters $c_1,c_2$ and $c_3$ given by (\ref{c1})-(\ref{c3})  and the properties of the CGLE allow us to verify that the system is always characterized by spiral waves in the coexistence phase. Therefore, while specific properties  (size, frequency and velocity) of the emerging spirals depend on the values $\sigma$ and $\mu$,
the general (qualitative) behavior of the system is already captured by the choice of parameters $\sigma=\mu=1$.

In the following, we describe how characteristic properties of the spiral waves, such as the spreading velocity, the selected frequency and wavelength can be inferred from the above CGLE (\ref{CGLE_eq}).
 
\subsection{The linear spreading velocity }
\label{subsec:velocity}

We have found [see, e.g., snapshots~(Fig.~\ref{snapshots})] that in the long-time regime the system exhibits traveling waves. 
Namely, in the steady state, regions with nearly only $A$ individuals are invaded by a front of $C$ individuals, which  is taken over by $B$ in turn,  and so on. The theory of front propagation into unstable states (see, e.g., \citep{saarloos-2003-386} and references therein) is useful to study analytically the related dynamics. Indeed, to determine the spreading velocity of the
propagating fronts one linearizes the CGLE~(\ref{CGLE_eq}) around the coexistence state $z=0$ [i.e. the reactive fixed point of~(\ref{ml_rate_eqs})], which yields
\begin{equation}
\partial_t z({\bm r},t)=D\Delta z({\bm r},t) + (c_1-i\omega)z({\bm r},t) +o(z^2)~.
\label{lin_z}
\end{equation}
It is then useful to perform a  Fourier transformation:
\begin{equation}
\tilde{z}({\bm k}, \Omega)=\int_{-\infty}^\infty d{\bm r}dt~z({\bm r},t)e^{-i{\bm k}.{\bm r}-i\Omega t}~,
\end{equation}
which, together with~(\ref{lin_z}), gives the following dispersion relation:
\begin{equation}
\Omega(k) = \omega+i(c_1-Dk^2)~,
\end{equation} 
where $k=|{\bm k}|$. As $\text{Im} \Omega(k)>0$ for $k^2<c_1/D$, the state $z=0$ is linearly unstable in this range of wavevectors $k$. This confirms the analysis of the spatially homogeneous RE~(\ref{ml_rate_eqs}), where we already found that the coexistence fixed point is unstable. As for other systems characterized by fronts propagating  into unstable states \citep{saarloos-2003-386}, from Eq.~(\ref{lin_z}) one can  now compute the linear spreading velocity, i.e. the speed $v^*$ at which fronts (e.g. generated by local perturbations around $z=0$) propagate.  Following a classic treatment, whose
 general derivation can, e.g., be found in \citep{saarloos-2003-386}, the spreading velocity is obtained by determining a wavevector $k_*$ according to 
\begin{equation}
\frac{d\Omega(k)}{dk}\bigg|_{k_*}=\frac{\text{Im}\Omega(k_*)}{\text{Im}k_*}\equiv v^*~.
\end{equation}
The first equality singles out $k_*$ and the second  defines the linear spreading velocity $v^*$.
Here, we find:
\begin{eqnarray}
\text{Re}k_*=0~, \quad \text{Im}k_*=\sqrt{c_1/D}~, \quad v^*=2\sqrt{c_1 D} ~.
\label{vel}
\end{eqnarray}
The comparison of this analytical prediction with numerical results has revealed a good agreement, as illustrated in
 Fig.~\ref{beta_velocity} and discussed in the corresponding previous section.

\subsection{Wavelength and frequency }
\label{subsec:wavelength}

To determine analytically the wavelength $\lambda$ and the frequency $\Omega$ of the spiral waves, 
 the (cubic) nonlinear terms of the CGLE~(\ref{CGLE_eq}) have to be taken into account. 
From the understanding gained in the previous sections, we make a traveling-wave ansatz 
$z({\bm r},t)=Ze^{-i\Omega t-i{\bm q}.{\bm r}}$  leading to the following dispersion relation (with $q=|{\bm q}|$)
\begin{equation}
\Omega( q)=\omega + i(c_1-Dq^2) -c_2(i +c_3)Z^2\,.
\end{equation}
Separating real and imaginary parts, we can solve for $Z$, resulting in
 $Z^2= (c_1-D q^2)/c_2$. As already found above, the range of wavevectors that yield traveling wave solutions is therefore given by $q<\sqrt{c_1/D}$. The dispersion relation can, upon eliminating $Z$, be  rewritten as
\begin{equation} 
\Omega(q)= \omega + c_3(Dq^2-c_1). 
\label{disp}
\end{equation}

As manifests on the RHS of (\ref{disp}), $\Omega$ comprises two contributions. On the one hand there is $\omega$, acting as a ``background frequency'', which stems from the nonlinear nature of the dynamics and  is already accounted by (\ref{ml_rate_eqs}) when the system is spatially homogeneous.  On the other hand, the second contribution on the RHS of (\ref{disp}) is  due to the spatially-extended character of the model and to the fact that traveling fronts propagate with velocity $v^*$,  therefore
generating oscillations with a frequency of $v^*q$. Both contributions superpose and, to sustain a velocity $v^*$, the dynamics selects a  wavenumber $q^\text{sel}$  according to
the relation $\Omega(q^\text{sel})=\omega+v^*q^\text{sel}$ ~\citep{saarloos-2003-386}. Solving this equation for $q^\text{sel}$ under the restriction $q^\text{sel}<\sqrt{c_1/D}$ yields 
\begin{equation}
q^\text{sel}=\frac{\sqrt{c_1}}{c_3\sqrt{D}}\left(1-\sqrt{1+c_3^2}\right)\,.
\label{qsel}
\end{equation}
Analytical expressions of the frequency $\Omega(q^\text{sel})$ and of the wavelength of the spirals, $\lambda=2\pi/q^\text{sel}$, can be obtained immediately from (\ref{disp}) and (\ref{qsel}). In fact, the frequency reads
\begin{equation}
\Omega= \Omega(q^\text{sel})=\omega+\frac{2c_1}{c_3}\left(1-\sqrt{1+c_3^2}\right)\,,
\label{omega}
\end{equation}
and the wavelength is given by
\begin{equation}
\lambda=\frac{2\pi c_3\sqrt{D}}{\sqrt{c_1}\big(1-\sqrt{1+c_3^2}\big)}\,.
\label{lambda}
\end{equation}

The expressions (\ref{qsel})-(\ref{lambda}) have been derived by considering a traveling wave ansatz as described above. The latter hold in \emph{arbitrary dimensions}. However, while traveling waves appear in one dimensions, in higher dimensions, the generic emerging structures are somewhat different. E.g. rotating spirals arise in two dimensions, as described in this article, while scroll waves are robust solutions of the CGLE (\ref{CGLE_eq}) in three spatial dimensions~\citep{aranson-2002-74}. However, the characteristic properties of these patterns, such as wavelength and frequency, still agree with those of traveling waves. Indeed, concerning the dynamical system investigated in this article, we have shown how the self-forming spirals are well characterized by the expressions~(\ref{omega}) and~(\ref{lambda}). The same system studied in three dimensions is therefore expected to exhibit an entanglement of scroll waves, whose wavelengths and frequencies are again given by Eqs.~(\ref{omega}) and~(\ref{lambda}).

\section{Discussion}

Individuals' mobility as well as intrinsic noise have crucial influence on the self-formation of spatial patterns. We have quantified their influence by investigating a stochastic spatial  model of mobile individuals experiencing cyclic dominance via interactions of  `rock-paper-scissors' type. 
We have demonstrated that individuals' mobility has drastic effects on the emergence of spatio-temporal patterns. Low exchange rate between neighboring individuals leads to the formation of small and irregular patterns. In this case coexistence of all subpopulations is preserved and the ensuing patterns are mainly determined by stochastic effects. On the other hand, in two dimensions, larger exchange rates (yet of same order as the reaction rates) yield the formation of  (relatively) regular spiral waves whose rotational nature is reminiscent of the cyclic and out-of-equilibrium ensuing kinetics. In fact, the three subpopulations endlessly, and in turn, hunt each other. The location and density of the spirals' vortices is either determined by initial spatial inhomogeneities, if these take pronounced shape, or by stochasticity. In the latter case, internal noise leads to an entanglement of many small spirals and a universal vortex density of about 0.5 per square wavelength. Increasing the diffusion rate (i.e. individuals' mobility), the typical size of the spiral waves rises, up to a critical value. When that threshold is reached, the spiral patterns outgrow the two-dimensional system and there is only one surviving subpopulation covering uniformly the system \citep{reichenbach-2007-448}.
 
The language of  interacting particles enabled us 
to devise a proper treatment of the  stochastic spatially-extended system
and to reach a comprehensive understanding of the resulting out-of-equilibrium and nonlinear phenomena. In particular, we have shown how  spatio-temporal properties of the system can be aptly described 
 in terms  of stochastic partial differential equations (SPDE) and confirmed our findings with lattice simulations. We have paid special attention to analyze the wavelength and frequency of the spiral waves, as well as the 
velocity of the  propagating fronts. Numerical solutions of the SPDE have been shown to 
share (statistically) the  same steady states
as the lattice simulations, with the emerging spiral waves characterized in both cases the same wavelength, overall sizes and frequency.
We have also studied the influence of stochasticity on the properties of the coexistence state and its spatio-temporal structure. 
Namely, we have compared the results obtained from the 
SPDE with those of the deterministic PDE (obtained by dropping the noise contributions  in the SPDE), which still yield spiralling structures. 
This allowed us to shed light on the fact that, in the presence of (sufficient) mobility, the wavelength and frequency of the spirals are not affected by internal noise.
However, there are major differences between the stochastic and deterministic descriptions of the system. One of the most important 
is the influence of the initial conditions. On the one hand, if initial spatial inhomogeneities are larger than the noise level, or if noise is absent as in the deterministic descriptions, these initial spatial structures determine the position of the spirals' vortices. In this situation, the system ``memorizes'' its initial state, and the latter crucially influences the overall size of the emerging spiral waves. On the other hand, for rather homogeneous initial densities (at values of the unstable reactive fixed point), the patterns emerging from the stochastic descriptions (lattice simulations and SPDE) are caused by noise and characterized by a universal density of $0.5$ spiral vortices per square wavelength. 
 While we have provided qualitative explanations of these findings,  a more profound understanding is still desirable
and could motivate further investigations. 

We have also shown that analytical expressions for the spirals' wavelength and frequency can be determined
by means of a complex Ginzburg-Landau equation (CGLE) obtained by recasting the PDE of the system, restricted onto
an invariant manifold, in a normal form. There is  good agreement between analytical predictions stemming from the system's CGLE
and the numerical results (obtained from stochastic lattice simulations as well as the numerical solution of the SPDE).
This can be traced back to the fact that
 May-Leonard rate equations are characterized by heteroclinic orbits very much reminiscent of 
limit cycles resulting from a Hopf bifurcation. 
The fact that the dynamics can be recast in the form of a  CGLE, known to give rise to the emergence of
coherent structures, reveals the generality of the phenomena discussed in this work and greatly facilitates their quantitative analysis. 
In particular, the emergence of an entanglement of spiral waves in the coexistence state, the dependence of spirals' size on the diffusion rate, and the existence of a critical value of the diffusion above which coexistence is lost are robust phenomena. 
This means that they do not depend on the details of the underlying spatial structure: While, for specificity,  we have (mostly) considered square lattices, other two-dimensional topologies (e.g. hexagonal or  other lattices) will lead
to the same phenomena, too. Also the details of the cyclic competition have
 no qualitative influence, as long as the underlying rate equations exhibit an unstable coexistence fixed point
and can be recast 
in the universality class of the Hopf bifurcations. 
We still note that instead of defining the model in terms of chemical reactions, as  done here~(\ref{ml_react}), 
we can equivalently  choose a formulation in terms of payoff matrices \citep{Smith,Hofbauer}.

We have investigated the system's behavior in two spatial dimensions. However, our approach, using a continuum limit to derive the SPDE~(\ref{stoch_part_eq}) as well as the CGLE~(\ref{CGLE_eq}), is equally valid in other dimensions and expected to describe the formation of spatial patterns, as long as the mobility is below a certain threshold value~\citep{reichenbach-2007-448}. As examples, in one dimension, the CGLE yields traveling waves, while ``scroll waves'', i.e. vortex filaments, result in three dimensions~\citep{aranson-2002-74}.

In this article, we have mainly focused on the situation where the exchange rate between individuals is sufficiently high, which leads to the emergence of regular spirals in two dimensions. However, when the exchange rate is low (or vanishes), we have seen that 
stochasticity strongly affects the structure of the ensuing spatial patterns. In this case, the (continuum) description in terms of SPDE breaks down. In this situation, the quantitative analysis of the spatio-temporal properties of interacting particle systems
requires the development of other analytical methods, e.g. relying on field theoretic techniques~\citep{mobilia-2007-128}. Fruitful insights into this regime have already been gained by pair approximations or larger-cluster approximations~\citep{tainaka-1994-50,sato-1997-47,szabo-2004-37,szabo-2007-446}. The authors of these studies investigated a set of  coupled nonlinear differential equations for the time evolution of the probability to find a cluster of certain size in a particular state. While such an  approximation improves when large clusters are considered, unfortunately the effort for solving their coupled equations of motion also drastically increases with the size of the clusters. 
In addition, the use of those cluster mean-field approaches becomes problematic in the proximity of phase transitions (near an extinction threshold) where the correlation length diverges.  Investigations along these lines represent a major future challenge in the multidisciplinary field of complexity science.

\section*{Acknowledgments}

Financial support of the German Excellence Initiative via the program ``Nanosystems
Initiative Munich" and the German Research Foundation via the SFB TR12 ``Symmetries and Universalities in Mesoscopic Systems''  is gratefully acknowledged.
M.~M. is grateful to the Alexander von Humboldt Foundation and to the Swiss National Science Foundation
for support through the grants IV-SCZ/1119205 and PA002-119487, respectively.

\begin{appendix}

\section*{Appendix A. Equations for the ${\bm y}$-variables}

In this Appendix, as well as the two following, some further details on the derivation of the normal form~(\ref{normal_form}) of the May-Leonard RE~(\ref{ml_rate_eqs}) are given. Namely, we derive the time evolution of the $y$-variables introduced in Subsection~\ref{subsec:IM}, and use the invariant manifold to eliminate the stable degree of freedom. The derivation of the invariant manifold is presented in the next section, while the transformation from  $y$- to  $z$-variables yielding the normal form is detailed in a third appendix.
  
In Subsection~\ref{subsec:IM}, we have introduced proper coordinates $(y_A,y_B,y_C)$ by 
 ${\bm y}=\mathcal{S}{\bm x}$ with the matrix $\mathcal{S}$ given by
\begin{equation}
\mathcal{S}=\frac{1}{3}\begin{pmatrix} \sqrt{3} & 0 & -\sqrt{3} \\
                                        -1 & 2 & -1 \\
					1 & 1 & 1          \end{pmatrix} \,.
\end{equation}
Hereby, the vector ${\bm x}=(a-a^*,b-b^*,c-c^*)^T$ encodes the deviation of the densities from the reactive fixed point. Of course, the time evolution of $\bm x$ is just given by the time evolution of the densities, Eqs.~(\ref{ml_rate_eqs}): $\partial_t {\bm x}=(\partial_ta,\partial_tb,\partial_tc)^T$. For the temporal evolution of the ${\bm y}$-variables, we have to apply the transformation given by $\mathcal{S}$: $\partial_t {\bm y}=\mathcal{S}\partial_t {\bm x}$. Expressing them in terms of $\bm y$-variables, we eventually obtain
\begin{align}
\partial_t y_A=&~\frac{\mu\sigma}{2(3\mu+\sigma)}\big[y_A+\sqrt{3}y_B\big] +\frac{\sqrt{3}}{4}\sigma\big[y_A^2-y_B^2\big]\cr
& \quad -\frac{\sigma}{2}y_Ay_B-\frac{1}{2}y_C\big[(6\mu+\sigma)y_A-\sqrt{3}\sigma y_B\big]\,,\cr
\partial_t y_B=&~\frac{\mu\sigma}{2(3\mu+\sigma)}\big[y_B-\sqrt{3}y_A\big] -\frac{\sigma}{4}\big[y_A^2-y_B^2\big]\cr
&-\frac{\sqrt{3}}{2}\sigma y_Ay_B-\frac{1}{2}y_C\big[\sqrt{3}\sigma y_A+(6\mu+\sigma)y_B\big]\,,\cr
\partial_ty_C=&~-\mu y_C-(3\mu+\sigma)y_C^2+\frac{\sigma}{4}\big[y_A^2+y_B^2\big]\,.
\label{eq_yaybyc}
\end{align}
Using the invariant manifold, Eq.~(\ref{center_eq}), we  eliminate $y_C$ from the above and are left with equations for $y_A,y_B$ alone. According to Eq.~(\ref{center_eq}) $y_C$ has been determined to second order in $y_A,y_B$, and as $y_C$ contributes to the time-evolution of $y_A,y_B$ through quadratic terms, we obtain $\partial_ty_A,\partial_ty_B$ up to third order:
\begin{align}
&\partial_t y_A=\frac{\mu\sigma}{2(3\mu+\sigma)}\big[y_A+\sqrt{3}y_B\big] +\frac{\sqrt{3}}{4}\sigma\big[y_A^2-y_B^2\big]-\frac{\sigma}{2}y_Ay_B\cr
&-\frac{\sigma(3\mu+\sigma)}{8\mu(3\mu+2\sigma)}\big(y_A^2+y_B^2\big)\big[(6\mu+\sigma)y_A-\sqrt{3}\sigma y_B\big] + o(y^3)\,,\cr
&\partial_t y_B=\frac{\mu\sigma}{2(3\mu+\sigma)}\big[y_B-\sqrt{3}y_A\big] -\frac{\sigma}{4}\big[y_A^2-y_B^2\big]-\frac{\sqrt{3}}{2}\sigma y_Ay_B\cr 
&\quad-\frac{\sigma(3\mu+\sigma)}{8\mu(3\mu+2\sigma}\big(y_A^2+y_B^2\big)\big[\sqrt{3}\sigma y_A+(6\mu+\sigma)y_B\big]+o(y^3)\,.
\label{eq_yayb}
\end{align}
These equations describe the system's temporal evolution on the invariant manifold.

\section*{Appendix B. The invariant manifold}

We provide further details concerning the derivation (to second order) of the invariant manifold of the RE~(\ref{ml_rate_eqs}) given by Eq.~(\ref{center_eq}).

To determine the invariant manifold parameterized by  $y_C=G(y_A,y_B)$ up to second order in $y_A,y_B$,
we make the ansatz $G(y_A,y_B)=K(y_A^2+y_B^2) + o(y^2)$, and calculate $K$. The condition~(\ref{eq_G}) turns into
\begin{equation}
2Ky_A\partial_t y_A+2Ky_B\partial_t y_B=\partial_t y_C\Big|_{y_C=G}~.
\end{equation}
Using Eqs.~(\ref{eq_yaybyc}), we obtain (to second order in $y_A,y_B$)
\begin{equation}
K\frac{\mu\sigma}{3\mu+\sigma}\big(y_A^2+y_B^2\big)=\Big(-\mu K+\frac{\sigma}{4}\Big)\big(y_A^2+y_B^2\big)~,
\end{equation}
which is satisfied for
\begin{equation}
K=\frac{\sigma}{4\mu}\frac{3\mu+\sigma}{3\mu+2\sigma}~.
\end{equation}

\section*{Appendix C. Normal Form: the nonlinear transformation ${\bm y}\rightarrow{\bm z}$}

The normal form~(\ref{normal_form}) of the May-Leonard RE~(\ref{ml_rate_eqs}) follows from the time-evolution equations in the $y$-variables, given by~(\ref{eq_yayb}), through an additional nonlinear variable transformation. Here, we present the latter.

The equations of motion~(\ref{eq_yayb}) comprise quadratic and cubic terms. 
To recast Eqs.~(\ref{eq_yayb}) in their normal form, we seek a transformation allowing to eliminate
 the quadratic terms. We make the ansatz of a quadratic transformation $\bm{y}\to \bm{z}$ and determine the coefficients
by cancelling the quadratic contributions to the RE in the $\bm{z}$ variables, this leads to
\begin{align}
z_A=&~y_A+\frac{3\mu+\sigma}{28\mu}[\sqrt{3}y_A^2+10y_Ay_B-\sqrt{3}y_B^2]\,,\cr
z_B=&~y_B+\frac{3\mu+\sigma}{28\mu}[5y_A^2-2\sqrt{3}y_Ay_B-5y_B^2]\,.
\label{transf_y_z}
\end{align}
To second order, this nonlinear transformation can be inverted: 
\begin{align}
y_A=&~z_A-\frac{3\mu+\sigma}{28\mu}[\sqrt{3}z_A^2+10z_Az_B-\sqrt{3}z_B^2]\cr &+\frac{(3\mu+\sigma)^2}{14\mu^2}[z_A^3+z_Az_B^2]+o(z^3)\,,\cr
y_B=&~z_B-\frac{3\mu+\sigma}{28\mu}[5z_A^2-2\sqrt{3}z_Az_B-5z_B^2]\cr
&+\frac{(3\mu+\sigma)^2}{14\mu^2}[z_A^2z_B+z_B^3]+o(z^3)\,.
\end{align}
With these expressions, one can check that equations of motion (\ref{eq_yayb}) are recast
in the normal form (\ref{normal_form}).
%

\section*{Appendix D. Kramers-Moyal expansion of the master equation}

For a  large number $N$ of interacting individuals, the master equation describing the stochastic system may be expanded in the system size $N$, often referred to as Kramers-Moyal expansion~\citep{VanKampen,Gardiner,Tauber}. As a result, one obtains a Fokker-Planck equation which is equivalent to a set of Ito stochastic differential equations (with white noise). In Subsection~\ref{subsec:noise}, we have shown that for the present stochastic spatial system, only noise terms stemming from the reactions~(\ref{ml_react}) contribute. The latter may be derived considering the stochastic non-spatial system. Here, we follow this approach. Starting from the master equation for the stochastic well-mixed system, we detail the system size expansion, and derive the Fokker-Planck equation as well as the corresponding Ito stochastic differential equations.

Denote ${\bm s}=(a,b,c)$ the frequencies of the three subpopulations $A$, $B$, and $C$. 
The Master equation for the time-evolution of the probability $P({\bm s},t)$ of finding the system in state ${\bm s}$ at time $t$ reads
\begin{align}
\partial_t P({\bm s},t)=\sum_{\delta{\bm s}}\big\{&P({\bm s}+\delta{\bm s},t)\mathcal{W}({\bm s}+\delta{\bm s}\rightarrow {\bm s})\cr
&- P({\bm s},t)\mathcal{W}({\bm s}\rightarrow {\bm s}+\delta{\bm s}) \big\}\,.
\label{master_eq}  
\end{align}
Hereby, $\mathcal{W}({\bm s}\rightarrow {\bm s}+\delta{\bm s})$ denotes the transition probability from state ${\bm s}$ to the state ${\bm s}+\delta{\bm s}$ within one time step; summation extends over all possible changes $\delta{\bm s}$. The relevant changes $\delta\bm s$ in the densities result from the basic reactions (\ref{ml_react}); as an example, concerning the change in the density of the subpopulation $A$, it reads $\delta s_A=1/N$ in the reaction $A\oslash\stackrel{\mu}{\longrightarrow} AA$,  $\delta s_A=-1/N$ in the reaction $CA\stackrel{\sigma}{\longrightarrow} C\oslash$, and zero in the remaining ones. Concerning the rates for these reactions, we choose the unit of time such
that, on average, every individual reacts once per time step. The transition rates resulting from the reactions~(\ref{ml_react}) then read  $\mathcal{W}=N\sigma ac$ for the reaction 
$CA\stackrel{\sigma}{\longrightarrow} C\oslash$
and $\mathcal{W}=N\mu a(1-a-b-c)$ for $A\oslash\stackrel{\mu}{\longrightarrow} AA$. Transition probabilities associated with all other reactions (\ref{ml_react}) follow analogously.

The  Kramers-Moyal expansion~\citep{Tauber} of the Master equation is an expansion in the increment $\delta{\bm s}$, which is proportional to $N^{-1}$. Therefore, it may be understood as an expansion in the inverse system size $N^{-1}$. To second order in $\delta{\bm s}$, it yields  the (generic) Fokker-Planck equation~\citep{Tauber}:
\begin{equation}
\partial_tP({\bm s},t)=-\partial_i[\alpha_i({\bm s})P({\bm s},t)]+\frac{1}{2}\partial_i\partial_j[\mathcal{B}_{ij}({\bm s})P({\bm s},t)] ~.
\label{fokker_planck}
\end{equation}
Hereby, the summation convention  implies sums carried over the indices $i,j\in \{A,B,C\}$. 
 According to the  Kramers-Moyal expansion, the quantities 
$\alpha_i$ and $\mathcal{B}_{ij}$ read~\citep{Tauber}
\begin{align}
\alpha_i({\bm s})=&\sum_{\delta{\bm s}} \delta s_i\mathcal{W}({\bm s}\rightarrow {\bm s}+\delta {\bm s})\,,  \cr
\mathcal{B}_{ij}({\bm s})=&\sum_{\delta {\bm s}}\delta s_i \delta s_j\mathcal{W}({\bm s}\rightarrow {\bm s}+\delta{\bm s}) \,.
\end{align}
Note that $\mathcal{B}$  is symmetric.
As an example, we now present the calculation of $\alpha_A({\bm s})$. The relevant changes $\delta s_A=\delta a$ result from the  reactions $A\oslash\stackrel{\mu}{\longrightarrow} AA$ and $CA\stackrel{\sigma}{\longrightarrow} C\oslash$. The corresponding rates as well as the changes in the density of subpopulation $A$ have been given above; together, we obtain $\alpha_A({\bm s})=\mu a(1-a-b-c)-\sigma ac$.
The other quantities are computed analogously; eventually, one finds 
\begin{align}
\alpha_A({\bm s})&=\mu a(1-a-b-c)-\sigma ac\,,\cr
\alpha_B({\bm s})&=\mu b(1-a-b-c)-\sigma ab \,,\cr
\alpha_C({\bm s})&=\mu c(1-a-b-c)-\sigma bc \,,
\end{align}
and 
\begin{align}
\mathcal{B}_{AA}({\bm s})&=N^{-1}\left[\mu a(1-a-b-c)+\sigma ac\right]\,,\cr
\mathcal{B}_{BB}({\bm s})&=N^{-1}\left[\mu b(1-a-b-c)+\sigma ab\right]\,,\cr
\mathcal{B}_{CC}({\bm s})&=N^{-1}\left[\mu c(1-a-b-c)+\sigma bc\right]\,.\cr
\end{align}  
The well-known correspondence between Fokker-Planck equations and Ito calculus~\citep{Gardiner} implies that~(\ref{fokker_planck}) is equivalent to the following set of Ito stochastic differential equations:
\begin{align}
\partial_t a&=\alpha_A + \mathcal{C}_{AA}\xi_A\,,\cr
\partial_t b&=\alpha_B + \mathcal{C}_{BB}\xi_B\,,\cr
\partial_t c&=\alpha_C  + \mathcal{C}_{CC}\xi_C \,.
\end{align}
Hereby,  the $\xi_i$ denotes (uncorrelated) Gaussian white noise terms. The matrix $\mathcal{C}$ is defined from $\mathcal{B}$
via the relation $\mathcal{C}\mathcal{C}^T=\mathcal{B}$~\citep{Gardiner}. As $\mathcal{B}$ is diagonal, we may choose $\mathcal{C}$ diagonal as well, with the square roots of the corresponding diagonal entries of $\mathcal{B}$ on the diagonal. We obtain the expressions~(\ref{C}).

\end{appendix}

{\small




}

\end{document}